\documentclass[aps,floats,floatfix,showpacs,amssymb,prd,twocolumn,superscriptaddress,nofootinbib,nolongbibliography,reprint,preprintnumbers]{revtex4-2}

\usepackage{array}[=2016-10-06]

\usepackage{amssymb,amsmath,verbatim,mathtools,needspace,enumitem,etoolbox,graphicx,physics,microtype,afterpage,xspace,tabularx,lmodern,multirow,bm,booktabs,nicefrac}
\usepackage{textcomp, gensymb}
\usepackage[algoruled]{algorithm2e}
\usepackage{aas_macros}

\usepackage{braket}

\usepackage{relsize}

\usepackage{dcolumn}

\usepackage[normalem]{ulem}
\usepackage{placeins}
\usepackage{comment}
\usepackage[dvipsnames]{xcolor}
\definecolor{linkcolor}{rgb}{0.0,0.3,0.5}
\usepackage[unicode, colorlinks=true, linkcolor=linkcolor, citecolor=linkcolor, filecolor=linkcolor, urlcolor=linkcolor, linktocpage, breaklinks]{hyperref}
\usepackage[all]{hypcap}
\usepackage[utf8]{inputenc}
\usepackage[T1]{fontenc}
\usepackage{mathrsfs}
\hypersetup{colorlinks=true,citecolor=romared,linkcolor=romared,urlcolor=romared}

\definecolor{romared}{RGB}{142,0,28}

\newcommand{\be}{\begin{equation}}
\newcommand{\ee}{\end{equation}}

\def\be{\begin{equation}}
\def\ee{\end{equation}}
\newcommand{\beq}{\begin{eqnarray}}
\newcommand{\eeq}{\end{eqnarray}}

\usepackage{aas_macros}
\usepackage{makecell}
\usepackage{soul}
\usepackage[nolist,nohyperlinks]{acronym}
\usepackage{orcidlink}
\usepackage{lipsum}

\acrodef{LSC}[LSC]{LIGO Scientific Collaboration}
\acrodef{BH}{black hole}
\acrodef{NS}{neutron star}
\acrodef{PN}{Post-Newtonian}
\acrodef{BBH}{binary black-hole}
\acrodef{BNS}{binary neutron-star}
\acrodef{NSBH}{neutron-star black-hole}
\acrodef{NR}{numerical relativity}
\acrodef{GW}{gravitational wave}
\acrodef{PSD}{power spectral density}
\acrodef{aLIGO}{Advanced Laser interferometer Gravitational-Wave Observatory}
\acrodef{AZDHP}{aLIGO zero detuned high power density}
\acrodef{GR}{general relativity}
\acrodef{PE}{parameter estimation}
\acrodef{LAL}{LIGO algorithm library}
\acrodef{TPI}{tensor-product interpolant}
\acrodef{SVD}{singular value decomposition}
\acrodef{SNR}{signal-to-noise ratio}
\acrodef{ODE}{ordinary differential equation}
\acrodef{PDE}{partial differential equation}
\acrodef{ROM}{reduced order model}
\acrodef{QNM}{quasi-normal mode}
\acrodef{IMR}{inspiral-merger-ringdown}
\acrodef{LVK}{LIGO-Virgo-KAGRA}
\acrodef{SXS}{Simulating eXtreme Spacetimes}

\newcommand{\jhu}{\affiliation{William H. Miller III Department of Physics and Astronomy,\\ Johns Hopkins University, 3400 North Charles Street, Baltimore, Maryland, 21218, USA}}
\newcommand{\UTAustin}{\affiliation{Weinberg Institute, University of Texas at Austin, Austin, TX 78712, USA}}
\newcommand{\uiuc}{\affiliation{Illinois Center for Advanced Studies of the Universe \& Department of Physics,\\University of Illinois Urbana-Champaign, Urbana, Illinois 61801, USA}}
\newcommand{\toulouse}{\affiliation{Universit\'{e} de Toulouse, CNRS/IN2P3, L2IT, Toulouse, France}}

\newcommand{\sissa}{\affiliation{SISSA, Via Bonomea 265, 34136 Trieste, Italy and INFN Sezione di Trieste}}

\graphicspath{{../images}}

\newcommand{\ben}{\begin{enumerate}}
\newcommand{\een}{\end{enumerate}}

\def\be{\begin{equation}}
\def\ee{\end{equation}}
\def\beq{\begin{eqnarray}}
\def\eeq{\end{eqnarray}}

\def\apjl{\rm{ApJ}}
\def\apjs{\rm{ApJS}}
\def\aap{\rm{A\&A}}

\defcitealias{Yi:2025pxe}{Paper~1}

\graphicspath{{./Plots/}}

\allowdisplaybreaks

\begin{document}

\pagenumbering{arabic}

\title{Systematic biases in parameter estimation on LISA binaries.\texorpdfstring{\\}{ }II. The effect of excluding higher harmonics for spin-aligned, high-mass binaries
}

\author{Sophia Yi\texorpdfstring{\,}{ }\orcidlink{0000-0002-9104-1734}}
\email{syi24@jh.edu}
\jhu

\author{Francesco Iacovelli\texorpdfstring{\,}{ }\orcidlink{0000-0002-4875-5862}}
\email{fiacovelli@jhu.edu}
\jhu

\author{Emanuele Berti\texorpdfstring{\,}{ }\orcidlink{0000-0003-0751-5130}}
\email{berti@jhu.edu}
\jhu

\author{Rohit S. Chandramouli\texorpdfstring{\,}{ }\orcidlink{0000-0001-5229-2752}}
\email{rchandra@sissa.it}
\sissa

\author{Sylvain Marsat\texorpdfstring{\,}{ }\orcidlink{0000-0001-9449-1071}}
\email{sylvain.marsat@l2it.in2p3.fr}
\toulouse

\author{Digvijay Wadekar\texorpdfstring{\,}{ }\orcidlink{0000-0002-2544-7533}}
\email{jay.wadekar@utexas.edu}
\jhu
\UTAustin

\author{Nicolás Yunes\texorpdfstring{\,}{ }\orcidlink{0000-0001-6147-1736}}
\email{nyunes@illinois.edu}
\uiuc

\date{\today}

\begin{abstract}
The Laser Interferometer Space Antenna (LISA) will observe massive black hole binaries (MBHBs) with astoundingly high signal-to-noise ratio, leaving parameter estimation with these signals susceptible to seemingly small waveform errors. Of particular concern for MBHBs are errors due to neglected higher-order modes. We extend Yi {\it et al.} [\href{https://doi.org/10.1103/3gs3-gmb4}{Phys. Rev. D \textbf{112}, 124063 (2025)}] to examine errors due to neglected higher-order modes for MBHBs with nonzero (aligned) progenitor spins and total mass up to $10^8\,M_\odot$. For these very massive systems, there can be regions of parameter space in which the $(\ell, |m|)=(2,\,2)$ modes are no longer dominant with respect to higher-order ones. We find that the extent of systematic bias can change significantly when varying the progenitor spins of the binary. We also find that for the heaviest, and therefore shortest, MBHB signals, slight systematic errors can cause severe misinference of the sky localization parameters. We propose an improved likelihood optimization scheme with respect to previous work as a way to predict these effects in a computationally efficient manner. 

\end{abstract}

\maketitle

\section{\label{sec:intro}Introduction}

The observation of massive black hole binaries (MBHBs) with the planned Laser Interferometer Space Antenna (LISA) will present us with gravitational wave (GW) signals that are orders of magnitude louder than we have ever observed with current ground-based instruments~\cite{LISA:2017pwj,LISA:2024hlh}. This potentially opens the door for exquisite tests of general relativity, cosmology, and astrophysics, the findings of which will greatly increase and possibly challenge our understanding of the Universe~\cite{LISA:2022kgy,LISAConsortiumWaveformWorkingGroup:2023arg,Klein:2015hvg,LISACosmologyWorkingGroup:2022jok,LISA:2022yao}. 
However, achieving reliable and accurate inference from such loud LISA observations requires exceptionally accurate waveforms~\cite{Cutler:2007mi,Lindblom:2008cm}. With a signal-to-noise ratio (SNR) that is very high compared to current ground-based GW observations, what could once be considered small errors in waveforms will become the limiting factor in inferring accurately the parameters of MBHBs and in testing cosmology and general relativity~\cite{Cutler:2007mi,Vallisneri:2012qq,Gair:2012nm,Yunes:2025xwp}.

Of the various effects expected to contribute to inaccurate waveforms leading to systematic errors, one of the most relevant for the analysis of MBHBs is that of neglected higher-order modes. To our knowledge, a thorough analysis of the effect of varying mode content on MBHB parameter estimation (PE) in the LISA context was first carried out in Ref.~\cite{Pitte:2023ltw}. Then, in Ref.~\cite{Yi:2025pxe} (henceforth, \citetalias{Yi:2025pxe}), some of the authors further explored the systematic biases due to neglecting higher-order modes in MBHB systems with total detector-frame mass up to $10^6\,M_\odot$. That study found that systematic errors can dominate parameter estimation in this regime, often being considerably larger than statistical errors. Indeed, the systematic biases are so large that they cannot be accurately predicted within the linear-signal approximation used in, e.g., Ref.~\cite{Cutler:2007mi}. However,~\citetalias{Yi:2025pxe} also found that one can nevertheless obtain a reasonable estimate of systematic biases in this regime by directly optimizing the likelihood.

In reality, we expect LISA to observe even more massive binaries than what was considered in~\citetalias{Yi:2025pxe}. Moreover, that initial analysis considered only nonspinning binaries for simplicity. In this work, we examine how systematic biases change and become more severe for heavier MBHBs, the signals of which are expected to lie in LISA's most sensitive frequency band and which will therefore have even higher SNRs than the systems considered in~\citetalias{Yi:2025pxe}. We also examine the effect of spin in the present work, although we still restrict our analysis to nonprecessing binaries. 

To summarize our main findings, we observe that the importance of higher-order modes does indeed increase further as we increase the total mass, pushing the merger to lower frequencies. However, the importance of a given mode is not simple to explain in terms of binary parameters or even individual mode SNR, due to many contributing effects and highly non-negligible cross terms between different angular components of the signal. Moreover, the relative importance of higher-order modes, and therefore the severity of biases due to neglecting these modes, also varies significantly when we change the progenitor spins. To investigate these effects, we introduce several improvements to the likelihood optimization technique utilized in~\citetalias{Yi:2025pxe}. For the most massive and shortest-lived signals, the waveform error introduced by a neglected higher-order mode can cause severe mislocalization of the MBHB.

The remainder of this paper is organized as follows. In Sec.~\ref{sec:spin_mode_ordering}, we briefly motivate the careful study of systematic biases, as we consider MBHBs with increasing total mass and spin magnitudes. In Sec.~\ref{sec:confirm_NM_higher_mass}, we examine the usefulness of the direct likelihood optimization method introduced in~\citetalias{Yi:2025pxe} in predicting biases throughout the expanded parameter space under consideration
and discuss some extensions. In Sec.~\ref{sec:spin_dependence} we study the spin dependence of the systematic biases. One of the additional challenges in obtaining accurate parameter estimation results, compared to the cases in~\citetalias{Yi:2025pxe}, is that the systematic biases can be so severe for these extremely high-SNR events that the maximum likelihood is located in a near-degenerate position in the sky with respect to the true point. This phenomenon is explored in Sec.~\ref{sec:degenerate_sky_pos}. In Sec.~\ref{sec:utility_of_dual_annealing}, we discuss how the direct likelihood optimization technique can be useful in handling multimodalities including, but not limited to, the ones discussed in Sec.~\ref{sec:degenerate_sky_pos}. In Sec.~\ref{sec:conclusions} we present our conclusions and possible directions for future work. To improve readability, some technical material is relegated to three Appendices.

\section{Ordering of subdominant modes}\label{sec:spin_mode_ordering}

\begin{figure}[tbp]
\centering
\includegraphics[width=0.45\textwidth]{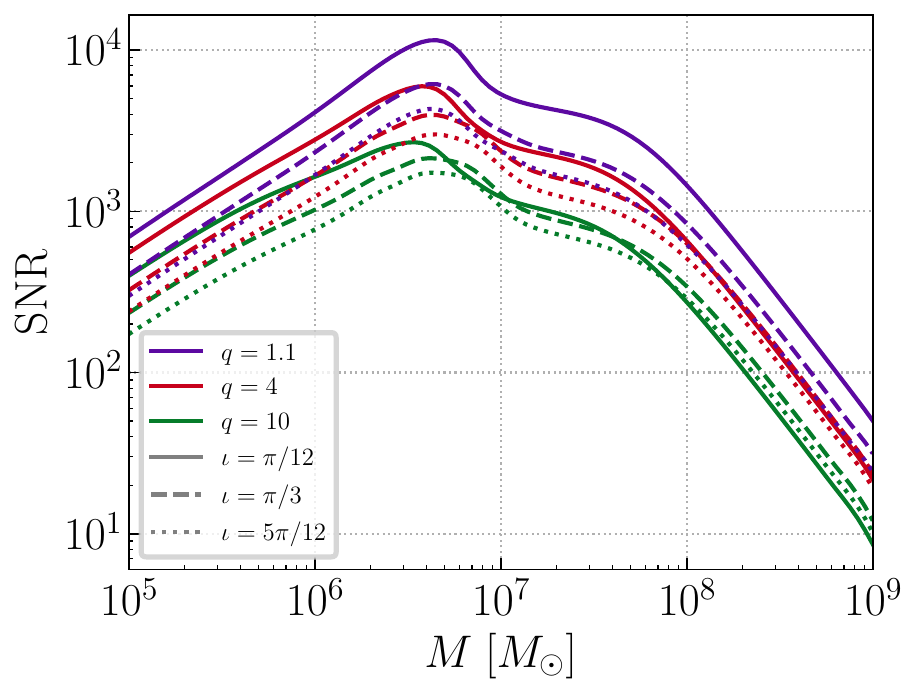} 
\caption{SNR vs. detector-frame total mass for a few values of mass ratio, $q$, and inclination angle, $\iota$, at redshift $z=1$ ($D_L=6791.8$\,Mpc). The lines are shown for nonspinning binaries ($\chi_1=\chi_2=0$). The remaining extrinsic parameters are the same as in~\citetalias{Yi:2025pxe} (namely, [$\delta t,\,\phi,\,\lambda_L,\,\beta_L,\,\Psi_L$] = [0,\,0.2,\,1.8,\,$\pi/6$,\,1.2]).%
}
\label{fig:SNR_vs_M}
\end{figure}

In this section, we demonstrate that the contribution of $\ell>2$ modes is extremely important for the most massive LISA sources, such that neglecting them in PE will lead to very inaccurate recovery of the parameters of the MBHBs. The loudest signals are expected for MBHBs with detector-frame mass between $10^6\,M_\odot$ and $10^7\,M_\odot$. This can be seen in Fig.~\ref{fig:SNR_vs_M}, where we plot the total SNR as a function of total detector-frame mass, $M$, for selected values of the mass ratio and inclination ($q=m_1/m_2>1$ and $\iota$, respectively, with $m_1$ and $m_2$ the primary and secondary component masses). We show the SNR at redshift $z=1$ (luminosity distance $D_L=6791.8\,{\rm Mpc}$). The SNR plotted here and referenced throughout this work is defined by $\rho^2=\left(h|h\right)$, with $h$ the full gravitational waveform and the inner product between two frequency-domain signals defined by
\begin{eqnarray}
   \label{eqn:inner_product}
    \left(a|b\right)=\sum_k\left(a_k|b_k\right)=4{\rm Re} \sum_k \int_0^\infty \mathrm{d}f \frac{a_k(f)b_k^*(f)}{S_{n,k}(f)}\,,
\end{eqnarray}
where the index $k$ in the sum refers to different time-delay interferometry (TDI) channels (e.g., $A,E,T$~\cite{Prince:2002hp}), the star denotes complex conjugation, and $S_{n,k}(f)$ is the one-sided noise power spectral density (PSD) of channel $k$. As in~\citetalias{Yi:2025pxe}, we use the PSD in the LISA Science Requirements Document (SciRDv1)~\cite{Babak:2021mhe} and the \textsc{IMRPhenomXHM} waveform model~\cite{Garcia-Quiros:2020qpx,Pratten:2020fqn} throughout, and we take the arbitrary parameter values [$\delta t,\,\phi,\,\lambda_L,\,\beta_L,\,\Psi_L$] = [0,\,0.2,\,1.8,\,$\pi/6$,\,1.2] for the time shift (in seconds) from coalescence time, the phase, the LISA-frame longitude and latitude, and the polarization angle, respectively.

From Fig.~\ref{fig:SNR_vs_M}, we also see that increasing either the mass ratio or the inclination angle (from 0 up to $\pi/2$) for a fixed total mass tends to decrease the system's SNR. These phenomena are consequences of the fact that a larger $q$ lowers the chirp mass and thus the amplitude, and moving from face-on to edge-on reduces the amplitude of the (2,\,2) %
signal component. To understand the latter, we recall that the $(\ell, m)$ component of the frequency-domain signal can be decomposed (using the stationary-phase approximation) as
\begin{eqnarray}
    h_{\ell m} = \frac{1}{D_L}\bar{h}_{\ell m} \,A_{\ell m} (\iota)e^{im\phi},
    \label{eqn:hlm_Alm}
\end{eqnarray}
where $D_L$ is the luminosity distance, and we use $\bar{h}_{\ell m}$ to denote the intrinsic amplitude of the $(\ell, m)$ mode, which depends only on the intrinsic parameters of the binary (masses and spins). The factor $A_{\ell m}(\iota)$ contains the inclination dependence obtained when projecting the signal onto the spin-weighted spherical harmonic basis; see, e.g., Ref.~\cite{Mills:2020thr} for explicit connections between $A_{\ell m}(\iota)e^{im\phi}$ and the spin-weighted spherical harmonics, $_{-2}Y_{\ell m}(\iota, \phi)$. In Fig.~1 of the same reference, the $A_{\ell m}(\iota)$ are plotted for the first few lowest-order harmonics, and the decrease of $A_{22}$ with respect to higher modes' geometric factors is evident as $\iota\rightarrow\pi/2$. We note that throughout this paper, we will often colloquially refer to the entire $(\ell, m)$ component of the signal defined in Eq.~\eqref{eqn:hlm_Alm} as a ``mode.'' However, when discussing the role of the geometric factor $A_{\ell m}(\iota)$ in suppressing or amplifying the signal, we will be more careful to refer to the full $(\ell,m)$ signal component defined in Eq.~\eqref{eqn:hlm_Alm} in order to avoid confusion with the intrinsic mode amplitude.

In this section, we study in detail the ordering of subdominant modes by SNR for MBHBs with total masses in the range $10^5-10^8\,M_\odot$. In Sec.~\ref{sec:degenerate_sky_pos}, we will examine interesting differences in the analyses of MBHBs with total masses $\sim\!10^5\,M_\odot$ vs. $\sim\!10^8\,M_\odot$. Although in Fig.~\ref{fig:SNR_vs_M} we observe that the total SNR of these two classes of signals is comparable, there are significant differences due to the differing length of the signal, and due to the fact that the (2,\,2) mode becomes less dominant with respect to higher-order modes as the total mass increases.

In Fig.~2 of~\citetalias{Yi:2025pxe}, the ordering of subdominant modes by SNR contribution was shown for MBHBs in the total mass range $3\times10^5<M/M_\odot<10^6$. While the ordering was found to depend considerably on the mass ratio and inclination angle, the (2,\,2) mode is always dominant in this regime. As we increase to even higher-mass MBHBs, %
we find that this is no longer necessarily the case. This is due to the fact that higher modes (particularly those with $m>2$) peak at higher frequencies compared to the $(2,2)$. As we consider more massive systems, which merge at lower frequencies, we can find ourselves in a regime in which the peak of the $(2,2)$ mode is near or below the lower end of LISA's sensitivity, while the peaks of higher modes (particularly the $(3,3)$ and $(4,4)$) remain in band. This frequency effect largely drives the hierarchy for $M \gtrsim 10^7\,M_\odot$. We also find in this section that allowing for nonzero progenitor spin also has a considerable effect on which modes contribute the most SNR in a given region of parameter space.  

\begin{figure*}[htbp]
\centering
\includegraphics[width=0.9\textwidth]{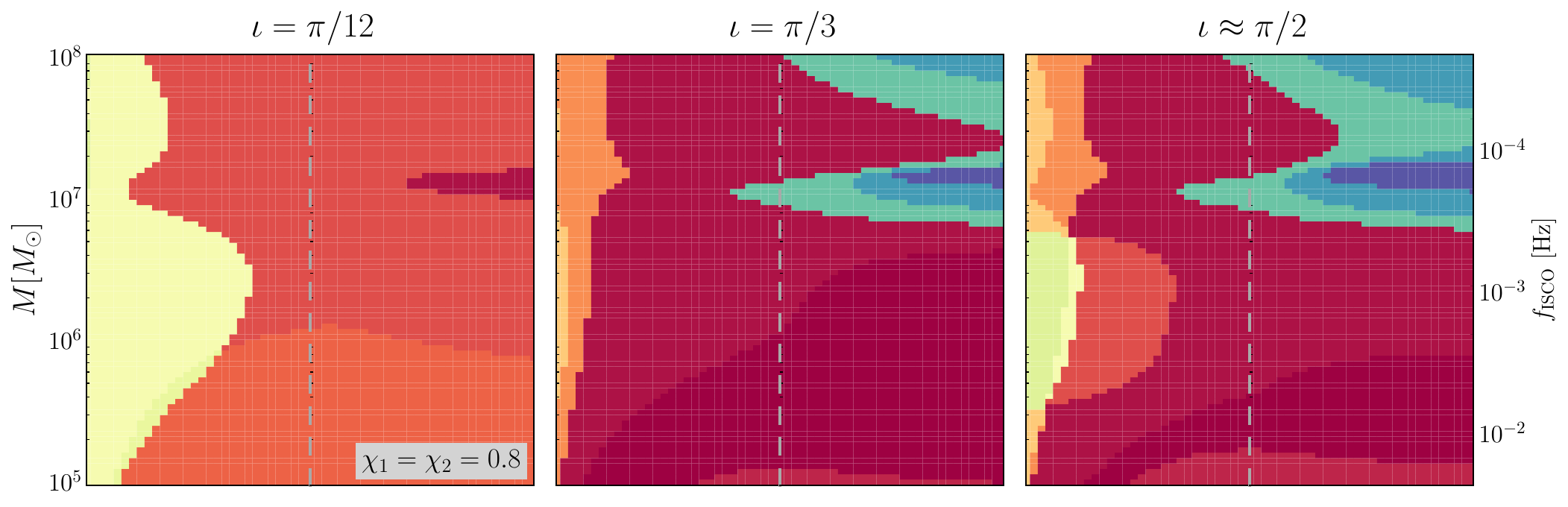} 
\includegraphics[width=0.9\textwidth]{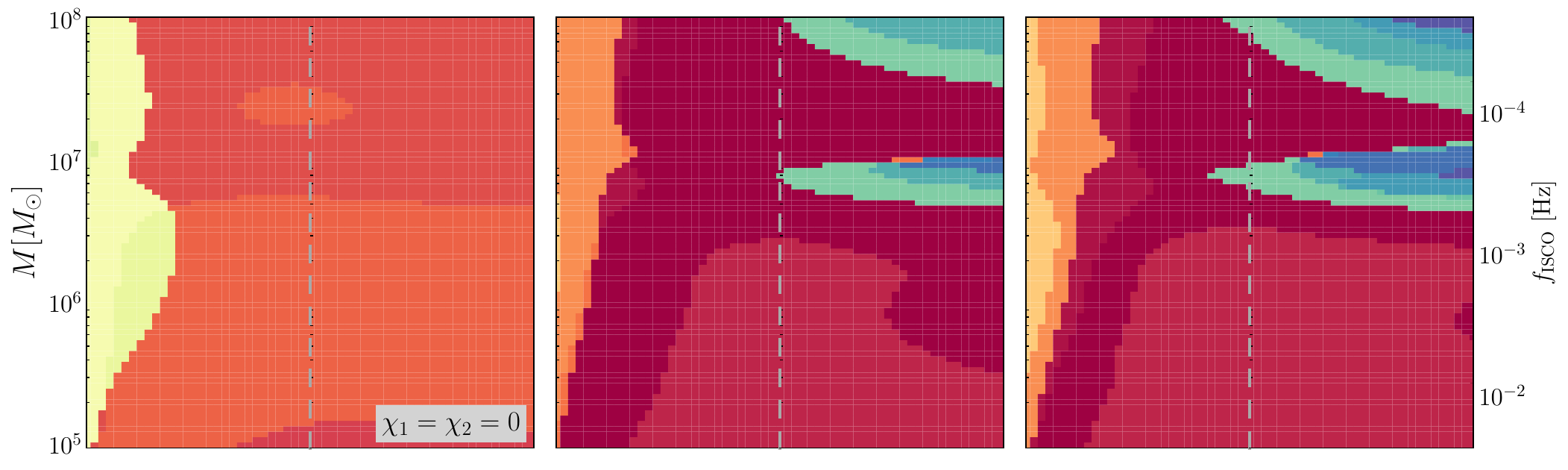} 
\includegraphics[width=0.9\textwidth]{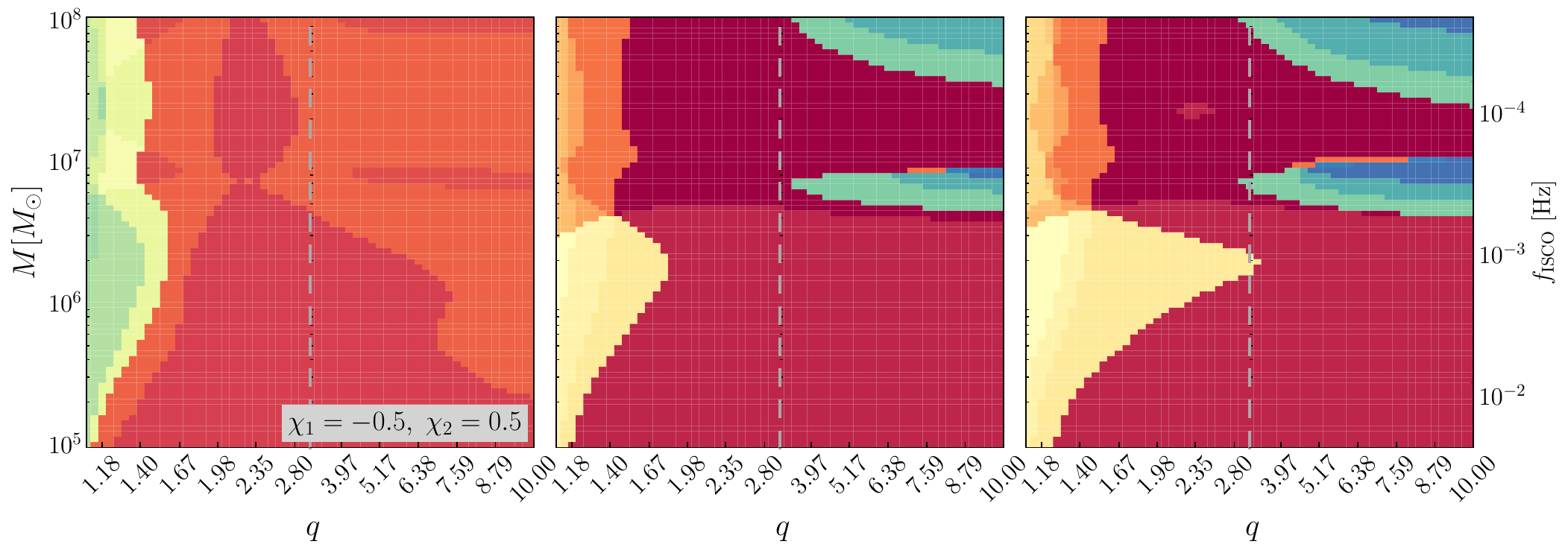} 
\hspace{12cm} \includegraphics[width=0.9\textwidth]{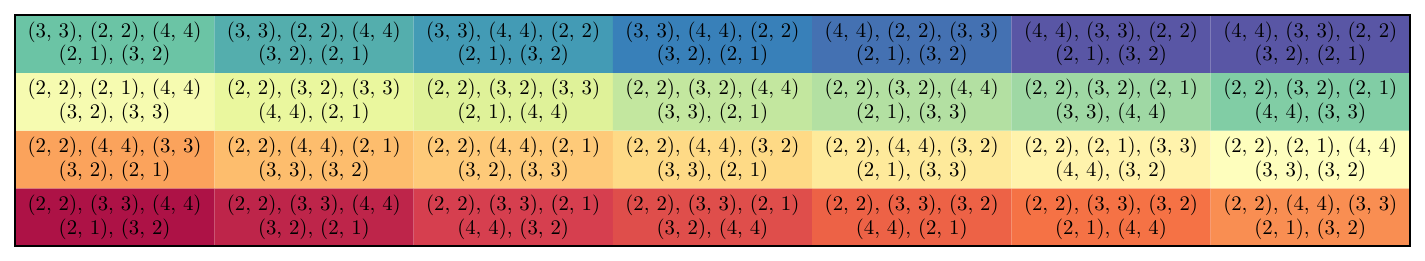} 
\caption{Hierarchy of SNR contribution by different $(\ell,m)$ signal components} as a function of total mass ($y$-axes), mass ratio ($x$-axes), inclination angle (column), and progenitor spins (row). As in~\citetalias{Yi:2025pxe}, 
we use log spacing between $q=1$ and $q=3$, as the importance of higher-order modes tends to change more rapidly between these mass ratios. The spacing is linear above $q=3$ (marked by the dashed gray line). Compared to~\citetalias{Yi:2025pxe}, the individual mode SNRs are ordered in many more different ways across this much broader region of parameter space including significantly higher masses and nonzero spins. Most notably, in the purple, blue, and darker green regions, the (2,\,2) mode contributes less SNR than the (3,\,3) and/or (4,\,4) modes.
Finally, the presence of a galactic white dwarf background causes nontrivial changes in the relative mode contribution around $M\sim10^7\,M_\odot$ ($10^{-4}\lesssim f_{\rm ISCO}/\rm{Hz}\lesssim 10^{-3}$). A comparison with Fig.~\ref{fig:mode_by_mode_noWD} of Appendix~\ref{app:frac_snr} clarifies that the features observed here are indeed due to the galactic background.
\label{fig:mode_by_mode}
\end{figure*}

We illustrate this dependence of the $(\ell,m)$ signal components on the total mass, mass ratio, inclination angle, and progenitor spins in Fig.~\ref{fig:mode_by_mode}. 
Particularly, we show the relative SNR contribution of the $(\ell,m) = (2,\,2),\, (3,\,3),\, (2,\,1),\, (4,\,4),$ and $(3,\,2)$ modes for binaries with total redshifted mass in the increased range of $10^5-10^8\,M_\odot$, mass ratio $q\in[1,\,10]$, and inclination angles $\iota=[\pi/12,\,\pi/3,\,5\pi/12]$ (shown in the left, center, and right columns, respectively). As in~\citetalias{Yi:2025pxe}, we take smaller increments between mass ratios 1 and 3, given the considerable changes in relative SNR contribution as we go from symmetric to asymmetric binaries. In the top row, we show the ordering (indicated by color) with both progenitors having high spins aligned with the orbital angular momentum of the binary; in the middle row, we show the ordering with zero spins; and in the bottom row, we show results for binaries having moderate progenitor spins completely antialigned with respect to one another (one antiparallel and one parallel to the orbital angular momentum). 

\begin{figure}[tbp]
\centering
\includegraphics[width=0.3727\textwidth]{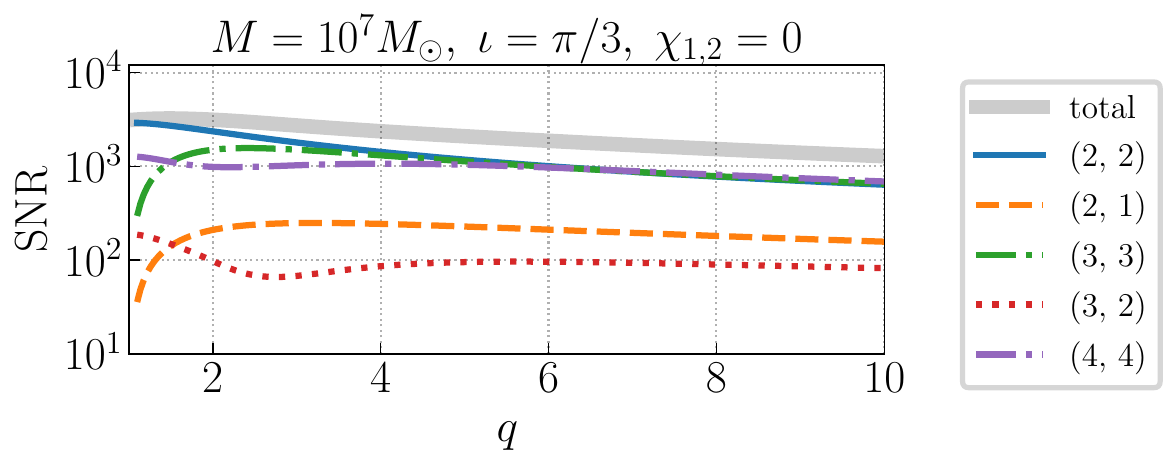}
\includegraphics[width=0.47\textwidth]{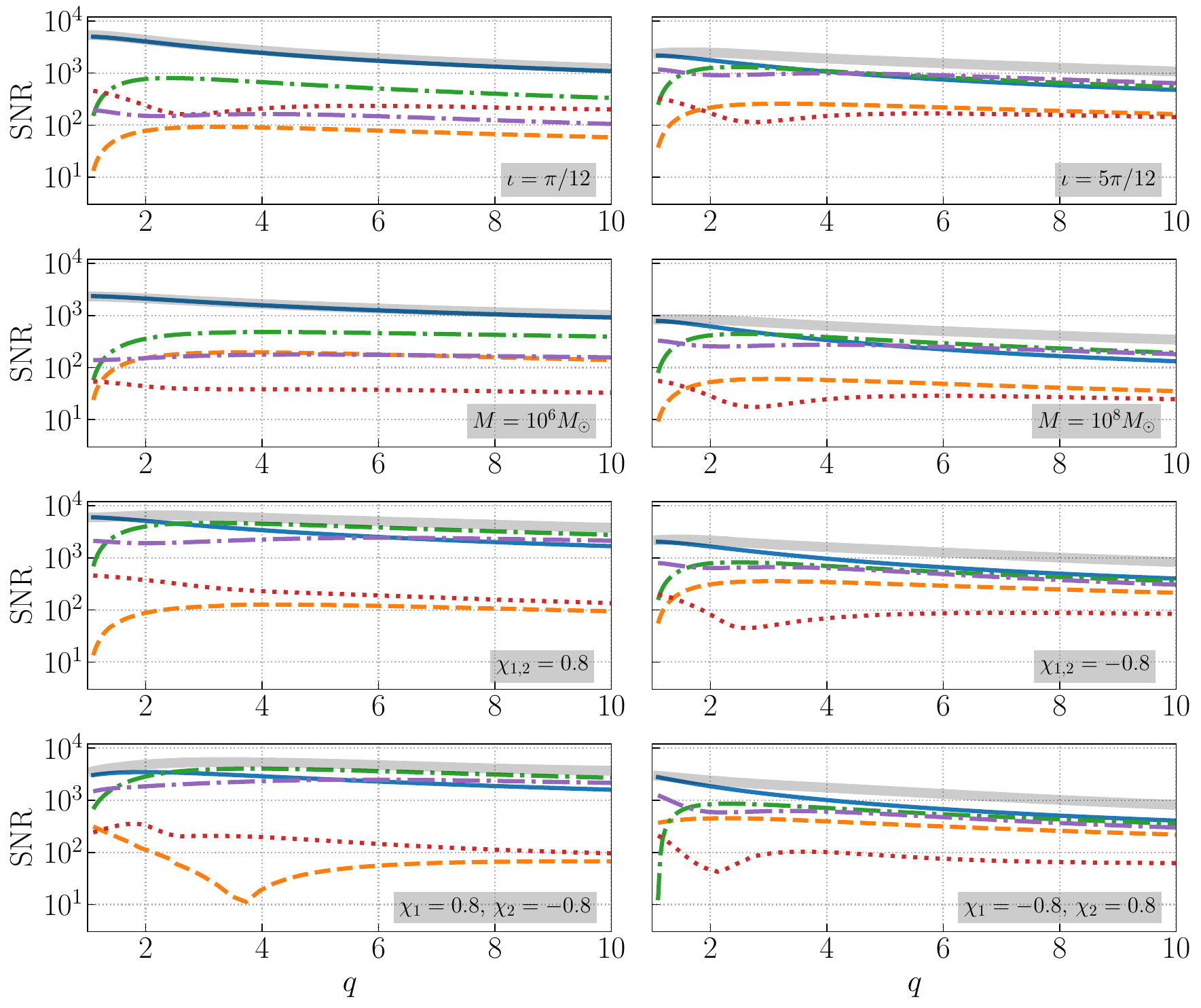} 
\caption{Hierarchy of SNR contribution by different $(\ell,m)$ signal components as a function of mass ratio. At the top, we show the hierarchy for a reference system of total mass $10^7\,M_\odot$, inclination $\pi/3$, and nonspinning progenitors. In the lower panels, we examine how this hierarchy changes as we modify one set of parameters at a time compared to the reference system (i.e., varying the inclination, total mass, and spins). In each panel, we show the total SNR in a thick light gray line. %
}
\label{fig:spin_dominance_condensed}
\end{figure}

\begin{figure}[tbp]
\centering
\includegraphics[width=0.47\textwidth]{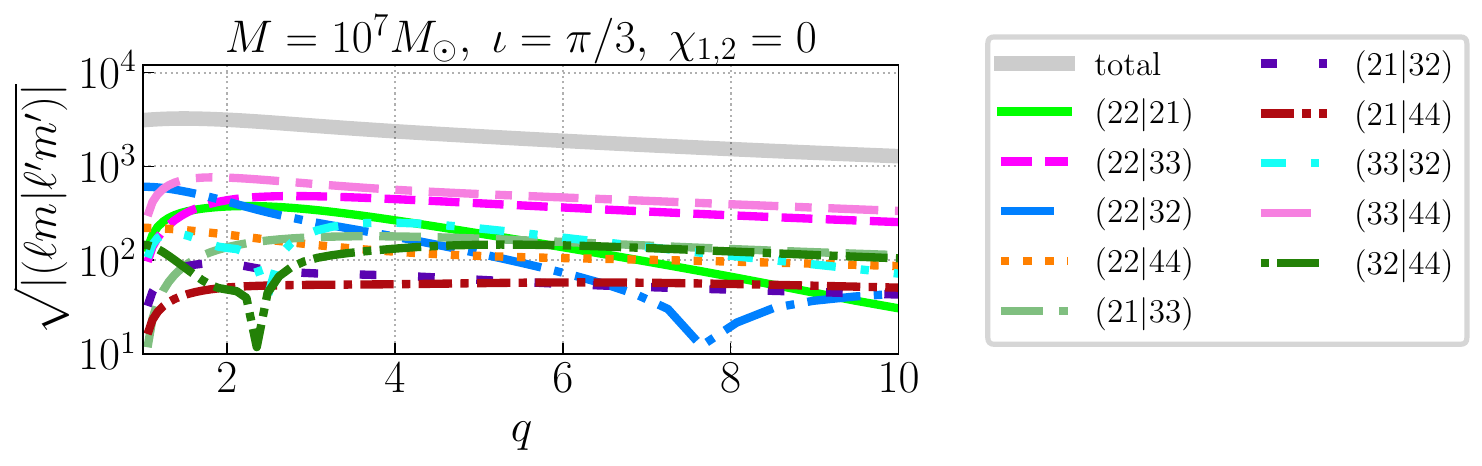}
\\
\includegraphics[width=0.47\textwidth]{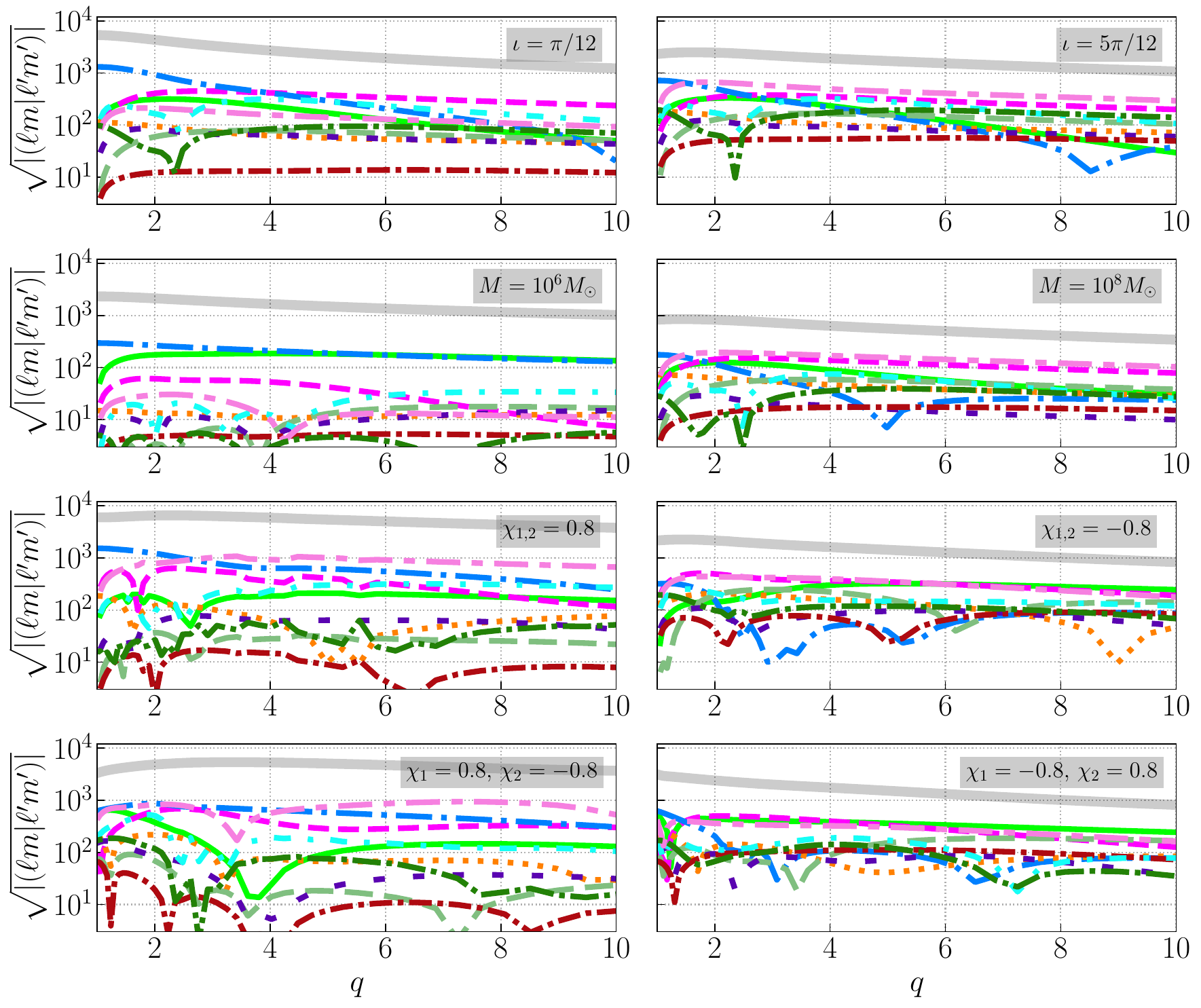} 
\caption{Same as Fig.~\ref{fig:spin_dominance_condensed}, but for the cross-term contributions to the signal. We again show the total SNR in a thick light gray line for reference. To improve readability, in this figure we denote $\left(h_{\ell m}|h_{\ell ' m'}\right)$ as simply $\left(\ell m|\ell'm'\right)$.
}
\label{fig:spin_dominance_condensed_cross}
\end{figure}

Overall, in Fig.~\ref{fig:mode_by_mode}, for relatively lower mass systems with $M \lesssim 10^6\,M_{\odot}$, we find that the $(2,\,2)$ mode is the most dominant.
However, there are regions of parameter space where the mode ordering is affected significantly by the total mass, spins, mass ratio, and inclination.
When the progenitor spins are zero, and with $\iota = \pi/12$, which is nearly face-on, we find that for all masses, the $(2,\,2)$ signal component is still the most dominant.
This is not surprising due to the fact that most higher angular factors ($A_{\ell m}$ in Eq.~\eqref{eqn:hlm_Alm}) have a power law dependence with $\sin \iota$.
However, the $(3,\,2)$ angular factor scales with the inclination as $\cos 2\iota$, which is why the $(3,\,2)$ signal component can sometimes (depending on masses) be higher in the ordering than, say, the $(3,\,3)$.
As the inclination increases, the $\{ (2,\,1),\, (3,\,3),\, (4,\,4) \}$ signal components get enhanced, while the $\{ (2,\,2),\, (3,\,2) \}$ components get suppressed, resulting in more regions (shaded by the bluer spectrum) of parameter space, depending on masses, where the quadrupolar $(2,\,2)$ component is no longer the most dominant one (as also noted in~\cite{Pitte:2023ltw}).

By extending to higher masses, even for zero spins, we see features that are more interesting when compared to~\citetalias{Yi:2025pxe}.
Further, as also noted in Ref.~\cite{Pitte:2023ltw}, even in regions where the $(2,\,2)$ multipole is loudest, it does not necessarily truly dominate the signal, in the sense that the fraction of SNR it contributes can be as little as $\sim$60\% (even less in cases where both progenitors have large negative spins).
In Appendix~\ref{app:frac_snr} we plot the fractional squared SNR contribution $\rho^2_{\ell m}/\rho^2$ of each mode for the $\iota=\pi/3,\;\chi_{1,2}=0$ case shown in the middle panel of Fig.~\ref{fig:mode_by_mode}. We define $\rho^2_{\ell m} = (h_{\ell m}|h_{\ell m})$ in terms of the inner product of Eq.~\eqref{eqn:inner_product}, where $h_{\ell m}$ is the waveform for a single mode with angular indices $\ell$ and $m$ (see Eq.~\eqref{eqn:hlm_Alm}).

Importantly, Figure~\ref{fig:mode_by_mode} shows that the presence of nonzero progenitor spins changes the ordering of subdominant modes considerably; the top, middle, and bottom rows of Fig.~\ref{fig:mode_by_mode} are noticeably distinct. There is a horizontal feature in the panels around $M\sim10^7\,M_\odot$ due to the presence of galactic binary confusion noise. This ``bump'' in the sensitivity alters the relative importance of the different $(\ell,m)$ signal components in nontrivial ways. On the right-most $y$-axes of Fig.~\ref{fig:mode_by_mode}, we plot the frequency of the dominant $(2,2)$ signal component at the innermost stable circular orbit (ISCO) for a Schwarzschild BH as a rough estimate for where the merger, and therefore loudest portion of the signal and largest contribution of higher-order modes, begins. When the MBHBs merge at frequencies near the range of significant galactic confusion noise, the ordering of the modes by SNR contribution is affected. In Fig.~\ref{fig:mode_by_mode_noWD} of Appendix~\ref{app:frac_snr}, we show how the mode ordering changes more smoothly without the presence of the confusion noise from galactic sources, with only the upper right corner of the parameter space corresponding to the $(2,\,2)$ mode no longer being the dominant one.

To see more precisely when the higher modes contribute greater SNR, as well as just how much more dominant they become compared to the (2,\,2) mode, in Fig.~\ref{fig:spin_dominance_condensed} we show the hierarchy of SNR in different $(\ell,m)$ components as a function of mass ratio, for a few fixed values of total mass, inclination angle, and spins. 
In these figures, it is clear that the contribution from odd-$m$ modes goes to zero in the equal-mass limit for non- or equally-spinning systems, as expected from symmetry arguments. The (2,\,2) component remains dominant at inclinations close to zero and when $M\leq10^6\,M_\odot$. For heavier binaries that are at least moderately away from a face-on orientation, the (3,\,3) and (4,\,4) signal components can rival the (2,\,2) in terms of SNR contribution. Even when the (2,\,2) is the loudest, it can clearly contribute a much lower fraction of the total SNR (thick gray line) at high mass ratios. For instance, at $q=10$ for the $\chi_{1}=\chi_2=-0.8$ case, the individual (2,\,2) SNR is only about 48\% of the total SNR. Finally, different spin configurations also alter the total and relative SNRs: a large, positive primary spin leads to a boost in the amplitudes of $\ell=m$ modes, while a large, negative spin leads to more equal weighting between the $\ell=m$ and $\ell\neq m$ modes. Note that the value of the secondary spin has comparatively little effect as the mass ratio increases (i.e., the bottom two rows look quite similar above $q\sim 5$).

For relatively low mass systems, which are dominated by their inspiral, one can obtain some understanding of the mode dependence on the mass ratio and spins using post-Newtonian (PN) theory
(see, e.g., Appendix~E of Ref.~\cite{Garcia-Quiros:2020qpx}).
The mass ratio enters the $(2,\,2)$-mode coefficient at 1\,PN order relative to the leading-order (LO) term. By contrast, it appears at 0.5\,PN order for the $\{(2,\,1),\,(3,\,3)\}$ modes and at 1\,PN order for the $\{(3,\,2),\,(4,\,4)\}$ modes, when measured relative to the LO $(2,\,2)$ contribution, which makes the former set of modes relatively more sensitive to the mass ratio. 
In Fig.~\ref{fig:mode_by_mode}, for $q \gtrsim 2$ and $M \lesssim 10^6\,M_{\odot}$, this PN-order dependence on the mass ratio explains why the $\{ (2,\,1),\, (3,\,3) \}$ modes are higher in SNR than the $\{ (3,\,2),\, (4,\,4) \}$ modes, provided the system is not nearly face-on --- reflected in the middle-center and middle-right panels with vanishing progenitor spins.
Further, the odd-$m$ modes are suppressed in the $q \rightarrow 1$ limit due to the dependence on $\delta = (m_1 - m_2)/(m_1 + m_2)$, which is clearly seen in Fig.~\ref{fig:spin_dominance_condensed}. 
At the leading PN order, there is no spin dependence in these modes, and the leading terms remain dominant when the spins are small, which is the case in the middle panels of Fig.~\ref{fig:mode_by_mode}.
The spin-dependent contributions enter (relative to the nonspinning contribution) at 0.5PN for odd $\ell + m$ modes (current source moments) and at 1.5PN for even $\ell + m$ modes (mass source moments).
Notably, relative to the leading nonspinning contribution for the $(2,\,2)$ mode, the spin-dependent terms enter at $\{1.5,\, 1,\, 2,\, 1.5,\, 2.5 \}$\,PN orders for the $\{ (2,\,2),\, (2,\,1),\, (3,\,3),\, (3,\,2),\, (4,\,4) \}$ modes, respectively. 
Further, whether the spins are aligned or antialigned has a strong influence on the ordering of the modes, irrespective of the inclination.
We can understand this by inspecting the coefficients of the leading spin contributions of each mode. For comparable masses, the leading spin contribution is governed by the antisymmetric aligned spin combination [$\chi_a = (\chi_{1,z} - \chi_{2,z})/2$] %
for the odd-$m$ ${(2,\,1),\,(3,\,3)}$ modes, and by the symmetric aligned spin combination [$\chi_s = (\chi_{1,z} + \chi_{2,z})/2$] for the even-$m$ ${(2,\,2),\,(3,\,2),\,(4,\,4)}$ modes, explaining the different ordering observed between the top and bottom rows of Fig.~\ref{fig:mode_by_mode}. This effect is most evident in the bottom right panel, which---compared to the top right panel---shows a larger region of parameter space with $q \lesssim 1.4$ where the $(2,\,1)$ mode dominates over the other higher modes. 

The progenitor spins further affect the characteristic merger frequency, as observed in numerical relativity simulations~\cite{Campanelli:2006uy} (see Ref.~\cite{McWilliams:2018ztb} for an analytical interpretation within the Backwards-One-Body model). For aligned (antialigned) spins, the merger frequency typically increases (decreases), thereby affecting whether each mode peaks nearer to or further from the sensitive part of the LISA band. Combined with the intrinsic spin dependence of the amplitudes, this leads to the observed spin-dependent ordering of the higher modes in the merger-dominated regime.

An important effect that is not captured in either Fig.~\ref{fig:mode_by_mode} or Fig.~\ref{fig:spin_dominance_condensed} is the impact of cross terms in the SNR. As explained in detail in Ref.~\cite{Pitte:2023ltw}, these cross terms contribute quite significantly (either positively or negatively) to the total signal, and are therefore also important to consider when investigating biases due to neglected modes. For instance, one can have a situation where $\rho^2_{21}$ is relatively small compared to the total squared SNR, but the cross term $\left(h_{22}|h_{21}\right)$ is quite sizable, leading to significant biases if the $(2,\,1)$ mode is excluded (see Fig.~3 of Ref.~\cite{Pitte:2023ltw}).
In general, as discussed in Ref.~\cite{Pitte:2023ltw}, the cross terms can oscillate between positive and negative values as the signal evolves in time or frequency, due to the difference in the phases between different modes, particularly when $m \neq m'$. The net accumulated contribution to the signal depends on both intrinsic and extrinsic parameters and is therefore not trivial to predict. In Fig.~\ref{fig:spin_dominance_condensed_cross}, we plot the square root of the absolute value of different cross terms for the same systems shown in Fig.~\ref{fig:spin_dominance_condensed}. 
For modes with $m \neq m'$, the corresponding cross terms are generally
strongly suppressed relative to the diagonal contributions, as the modes are out of phase. By contrast, for the $(h_{22}|h_{32})$ cross term, where $m = m'$, the modes are largely in phase through the inspiral, resulting in a %
$(h_{22}|h_{32})$ cross term that is often larger than its diagonal counterpart. %
Reference~\cite{Mills:2020thr} also examines the impact of cross-terms in the signal, quantifying effects in terms of the overlap. Given the similar frequency evolution between $m=m'$ modes in the inspiral and between $\ell=\ell'$ modes in the ringdown, one generally has larger overlaps between the $(2,2)$ and $(3,2)$ (inspiral contribution) or between the $(2,2)$ and $(2,1)$ (ringdown contribution). Neglecting these modes in parameter recovery consequently results in misattribution of power in the $(2,2)$ mode, potentially causing biases in the parameters that are largely determined by the quadrupole. In contrast, the approximate ``orthogonality'' of (smaller overlap with) the $(3,3)$ and $(4,4)$ modes is such that neglecting these modes is more likely to affect information in the signal that is not dominated by observation of the quadrupole.

Finally, we note that, as in~\citetalias{Yi:2025pxe}, we have not shown how the biases due to a neglected mode depend on the remaining extrinsic parameters $[\delta t, \, \phi,\,\lambda_L,\,\beta_L,\Psi_L]$. Of these, we have found that varying the azimuthal angle to the observer, $\phi$, can also have a moderate effect on the degree of systematic bias. We briefly illustrate this in App.~\ref{app:vary_phi}, and we note that a more extensive study of the effect of $\phi$, as well as other extrinsic parameters, could be an area for future work.

\section{Direct likelihood optimization for very massive and spinning binaries}\label{sec:confirm_NM_higher_mass}

\begin{figure*}[htb]
\centering
\includegraphics[width=0.99
\textwidth]{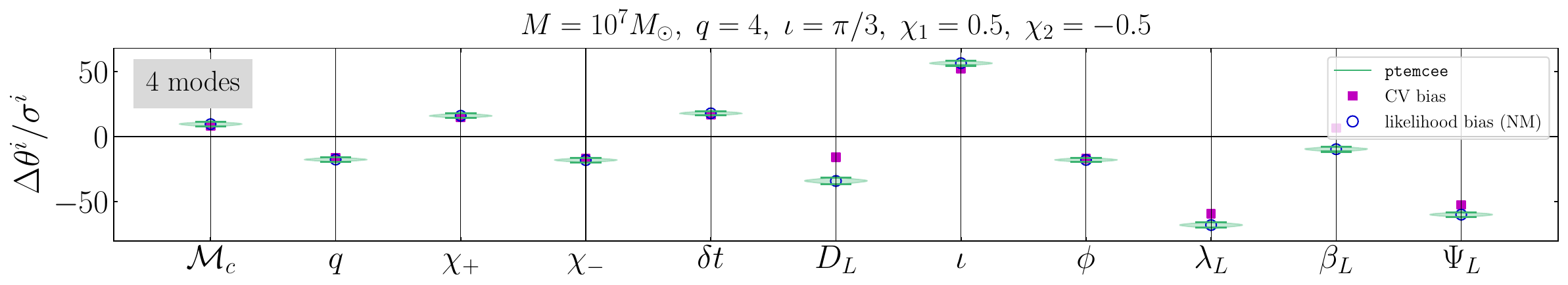} 
\includegraphics[width=0.98\textwidth]{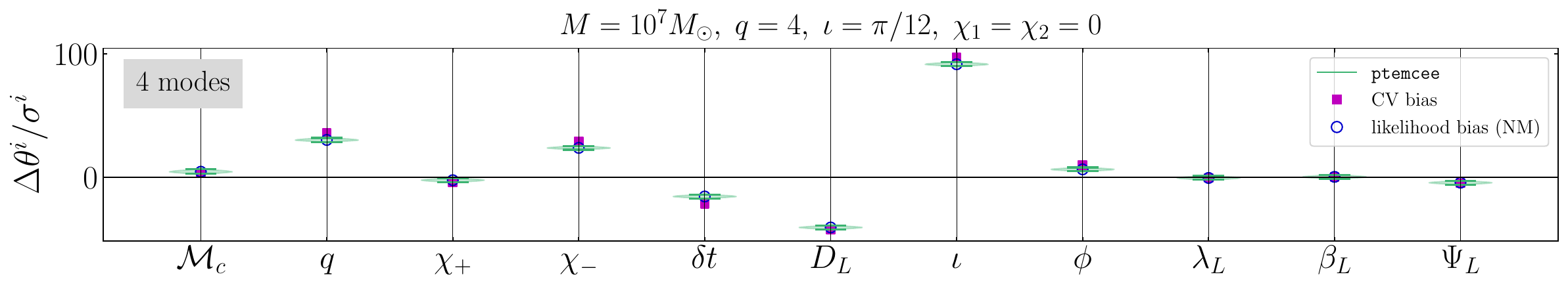} 
\caption{Comparison of the systematic biases on two MBHB parameters with 4 modes $=\{(2,\,2),\,(2,\,1),\,(3,\,3),\,(4,\,4)\}$ (excluding the $(3,\,2)$ mode), with biases determined by a full PE run (green violin plots and quantiles), the Cutler-Vallisneri approach (magenta squares), and the original likelihood optimization procedure used in~\citetalias{Yi:2025pxe} (blue circles labeled ``NM'' for the Nelder-Mead algorithm). The systematic biases are normalized by the respective statistical errors on each parameter.
In these cases, simple likelihood optimization via the Nelder-Mead algorithm continues to work well in estimating the systematic biases due to excluding higher-order modes for heavier events than were considered in Ref.~\cite{Yi:2025pxe}. Note that one of the systems considered here has nonzero aligned progenitor spins ($\chi_1=0.5,\chi_2=-0.5$).}
\label{fig:cv_nm_1e7}
\end{figure*}

In this section, we confirm that the direct likelihood optimization method utilized in~\citetalias{Yi:2025pxe} is still successful at estimating biases due to neglected higher modes for more massive events than were considered in the previous work, as well as for MBHBs with nonzero progenitor spins. In Sec.~\ref{sec:just_NM}, we discuss the cases in which likelihood optimization via the Nelder-Mead simplex algorithm~\cite{Nelder:1965zz} continues to work well, as in~\citetalias{Yi:2025pxe}. Then, in Sec.~\ref{sec:dual_annealing}, we discuss how various improvements (a global optimization technique, reparametrization, etc.) can be incorporated to have a more robust performance in the cases where a simple use of the Nelder-Mead algorithm fails. As in~\citetalias{Yi:2025pxe}, being interested in studying systematic biases due to waveform mismodeling, we always consider zero-noise injections. %

\subsection{Likelihood optimization with the Nelder-Mead simplex algorithm}\label{sec:just_NM}

In this section, we show that for the more severe biases that arise with the higher-SNR events having total mass $>10^6\,M_\odot$, direct likelihood optimization via the Nelder-Mead algorithm continues to work reasonably well long after the validity of the linear signal approximation has broken down. We remind the reader that the Nelder-Mead algorithm is a gradient-free optimizer in which a simplex with $n+1$ vertices is introduced in the $n$-dimensional space and methodically expanded, shrunk, flipped, and otherwise made to eventually collapse to the optimal point in parameter space.

In the two panels shown in Fig.~\ref{fig:cv_nm_1e7}, we show that directly maximizing the likelihood provides a bias estimate (blue circles) that recovers the bias in a full Bayesian PE (green violin plots), while the Cutler-Vallisneri (linear signal approximation) estimates fail to recover the results from PE in some cases (magenta squares). As in~\citetalias{Yi:2025pxe}, we use \texttt{lisabeta}~\cite{Marsat:2020rtl} with the \textsc{IMRPhenomXHM} waveform family~\cite{Garcia-Quiros:2020qpx,Pratten:2020fqn} and \texttt{ptemcee} sampler~\cite{Vousden:2016eeu,Foreman-Mackey:2012any} for PE. In addition to having a significantly higher total mass than what was considered in the previous work, one of these systems also has nonzero progenitor spins (in contrast with the previous work in~\citetalias{Yi:2025pxe}, which only considered nonspinning binaries).

The Cutler-Vallisneri / linear signal approximation for the systematic error is, by construction, a local estimator for the maximum likelihood. 
From our analysis, we find that the direct likelihood optimization methods are also only generally successful as local estimators of the maximum likelihood, and thus of the systematic bias.  
Specifically, the direct likelihood optimization methods are susceptible to finding large secondary maxima in the likelihood corresponding to near-degeneracies in the sky position. These degeneracies are the topic of Sec.~\ref{sec:degenerate_sky_pos} below.

A target often used in assessing systematic biases is requiring that the systematic errors be small enough that the resulting log-likelihood difference,
$\Delta \ln \mathcal{L}$, between the true waveform and the estimated maximum-likelihood waveform remains below fluctuations expected from an unknown noise realization.
A natural threshold is provided by the $p$-quantile of the log-likelihood difference distribution. 
In the high-SNR limit, Wilks' theorem implies that the quantity $2\,\Delta \ln \mathcal{L}$ is asymptotically distributed as $\chi^2_k$ \cite{Wilks1938,McWilliams:2010eq,Baird:2012cu,Toubiana:2024car}.
Accordingly, if
\begin{align}
   \Delta \ln \mathcal{L} \;< \; \dfrac{1}{2}\,\chi^2_k(p)\,,
   \label{eqn:logL_criterion}
\end{align}
then the systematic contribution to the log-likelihood difference lies within the $p$ fraction of the noise-induced distribution.
If the observed $\Delta\ln\mathcal{L}$ exceeds the right-hand side of Eq.~\eqref{eqn:logL_criterion}, then the log-likelihood difference is large enough to be distinguishable from random unknown noise fluctuations at $p\%$ confidence, i.e., the probability that such a large (or larger) log-likelihood difference would arise from noise alone is at most $1-p$ (the frequentist upper-tail probability). 

In our application, taking $k=11$ parameters and choosing $p=0.95\ (0.68)$ yields
$\Delta \ln \mathcal{L} \lesssim 9.84\ (6.30)$. %
We note that, strictly speaking, the criterion above applies only to Gaussian likelihoods. However, as the eightfold multimodality in the sky localization is nearly exact (and indeed, is quite exact when neglecting the time and frequency dependence of the LISA response~\cite{Marsat:2020rtl,Baibhav:2020tma}), Eq.~\eqref{eqn:logL_criterion} can still be adapted to describe the log-likelihood distribution of the multivariate Gaussian posterior. Nevertheless, in Sec.~\ref{sec:degenerate_sky_pos}, we will see that for very massive MBHBs, smaller changes in the likelihood than allowed by Eq.~\eqref{eqn:logL_criterion} can result in significant changes to the reconstructed physical parameters. One should therefore still be cautious in relying on criteria such as the one in Eq.~\eqref{eqn:logL_criterion} in the regime of very massive MBHBs, as it does not necessarily capture the fact that points with similar log-likelihood can correspond to considerably different extrinsic parameters. For more general non-Gaussian distributions, this criterion would not be valid at all.

\begin{algorithm}
    \DontPrintSemicolon
    \SetKwComment{Comment}{\#}{ }
    \SetKwInput{KwData}{Input}
    \SetKwInput{KwResult}{Output}
    \SetKwFunction{Fisher}{$\Gamma$}           
    \SetKwFunction{Cov}{${\cal C}$}           
    \SetKwFunction{Jac}{$J_r$}                  
    \SetKwFor{For}{for}{do}{}
    \SetKwFor{ForEach}{for each}{do}{}
    \SetKwFor{While}{while}{do}{}
    \SetKwIF{If}{ElseIf}{Else}{if}{then}{else if}{else}{}

    \KwData{Event parameters $\bm\theta_e$ for all the events ${\cal E}$; Map $r:\bm{\theta}\to\tilde{\bm{\theta}}$; Jacobians $\Jac$; Physical priors $\pi$; Restarts $N_{\rm res}$}
    \KwResult{Maximum-likelihood value ${\rm ln}{\cal L}_{{\rm max}}^{(e)}$ and parameters $\bm\theta^\star_e$ for each event $e \in \mathcal{E}$; Covariance matrices ${\cal C}_e$ for each event $e \in \mathcal{E}$}
    \ForEach{$e \in {\cal E}$}{
        $\tilde{\bm{\theta}}_e \gets \bm\theta_e$ \Comment*[r]{Parameters in Eq.~\eqref{eq:parameters_rep}}
        $\Cov_e\gets\Fisher_{e}^{-1} = [\Fisher(\bm{\theta}_e)]^{-1}$ \Comment*[r]{Fisher in Eq.~\eqref{eq:fisher_def}}
        $\Cov_e'\gets \Jac(\bm{\theta}_e)\, \Cov_e\, \Jac^{T}(\bm{\theta}_e)$\;
        $\sigma_{\tilde{\theta}} \gets [\Cov'_{e}]_{\tilde{\theta}\tilde{\theta}}^{1/2}$\;
        $\tilde{\bm{\theta}}_e \gets \tilde{\bm{\theta}}_e/\sigma_{\tilde{\theta}}$ \Comment*[r]{Rescaling}
        ${\rm lb}_{\tilde{\theta}}\gets {\rm max}(\pi_{\tilde{\theta}}^{\rm low}, \tilde{\theta}_e - 300\sigma_{\tilde{\theta}})$ \Comment*[r]{Bounds\,from\,Fisher}
        ${\rm ub}_{\tilde{\theta}} \gets {\rm min}(\pi_{\tilde{\theta}}^{\rm high}, \tilde{\theta}_e + 300\sigma_{\tilde{\theta}})$\;
        \For{$n = 1$ \KwTo $N_{\rm res}$}{
        	$\tilde{\bm\theta}_{{\rm init}, n}^{(e)} \gets {\cal N}(\tilde{\bm\theta}_e; \Cov_e')$ \Comment*[r]{Vary initialization}
        	$-{\rm ln}{\cal L}_{{\rm max}, n}^{(e)}, \tilde{\bm\theta}^{\star}_{e, n}\!\!\gets \!\texttt{minimize}(-{\rm ln}{\cal L}; \tilde{\bm\theta}_{{\rm init}, n}^{(e)}; {\rm lb}, {\rm ub})$\!\;
            \eIf{${\rm len}[{\rm all}(\Delta\tilde{\bm\theta}^{\star}_{e, n}\leq 10\%)] \geq 2$}{
                ${\rm ln}{\cal L}_{{\rm max}}^{(e)} \gets {\rm max}_n {\rm ln}{\cal L}_{{\rm max}, n}^{(e)}$\;
                $\bm\theta^\star_e \gets \tilde{\bm\theta}^\star_e \gets \tilde{\bm\theta}^{\star (e)}_{e, n} ({\rm ln}{\cal L}_{{\rm max}}^{(e)})$ \Comment*[r]{Inverse\,Eq.\,\eqref{eq:parameters_rep}}
            }{
                \texttt{NaN}
            }
        }
    }
    \caption{Workflow of the maximization procedure used in this work.}\label{alg:maximization}
\end{algorithm}

\subsection{Improvements to simple likelihood optimization}\label{sec:dual_annealing}

In some cases, we find that a basic implementation of the Nelder-Mead algorithm is not sufficient for predicting the systematic biases due to neglected higher-order modes. In such cases, we find it useful to utilize a global optimization tool called ``dual annealing,'' which is available from \texttt{scipy}~\cite{2020SciPy-NMeth}; to reparametrize the problem using a basis of parameters which are less correlated and degenerate among themselves; and to use priors informed from a Fisher analysis. A workflow of our implementation is reported in Algorithm~\ref{alg:maximization}, which we expand on below.

\subsubsection{Dual annealing}

The multidimensional, many-peaked nature of GW likelihoods leaves simple optimization techniques susceptible to finding incorrect, local maxima. Some multimodalities, such as those due to near-degeneracies in the sky localization, are anticipated ahead of time given our knowledge of the detector response, and can therefore be addressed in a straightforward manner. In Sec.~\ref{sec:degenerate_sky_pos}, we demonstrate how one can address this kind of multimodality by strategically restricting parameter inference to one region in the sky at a time. Even after addressing these major, anticipated multimodalities, however, the GW likelihood is still many-peaked in nature, and there can be additional multimodalities due to issues such as unforeseen systematics. While this was not found to be too extreme of an issue in~\citetalias{Yi:2025pxe}, it can indeed prevent us from finding the correct amount of systematic bias due to neglected higher-order modes in the regime of more massive events explored in the present work. 

To help prevent this effect, we utilize the \texttt{dual\_annealing} routine available from \texttt{scipy}, which occasionally allows ``downward'' steps in a maximization process based on a notion of ``temperature'' which is strategically decreased over time~\cite{Xiang:1997ycj,Tsallis:1987eu,1996PhyA..233..395T,PhysRevE.62.4473}. This allows the optimizer to escape local maxima and efficiently explore the areas around several ``peaks,'' thus enhancing the probability of eventually finding the highest peak. Importantly, the \texttt{dual\_annealing} routine allows for multiple different kinds of local optimizers. We continue to use the Nelder-Mead algorithm as the local optimizer, allowing the \texttt{dual\_annealing} part to determine where and for how long it is used in any given region.

There are several controls for the \texttt{dual\_annealing} routine that adapt it to different problems of interest. After testing various values for controls such as the starting ``temperature,'' the rate of ``re-annealing,'' etc., we generally find the default settings for most of these controls to work well for our purpose, although we leave a more thorough exploration of the effects of these parameters to future study. 

In general, there can be a small trade-off between greater stability and computation time when comparing the Nelder-Mead$+$dual annealing approach over simple Nelder-Mead. Nevertheless, 1000 iterations with dual annealing included still only take around 20-30 seconds on a single core (on average), compared to $\sim\!5$ minutes for a full PE run when parallelized across 48 cores (3.9\,GB RAM per CPU). In Sec.~\ref{sec:utility_of_dual_annealing}, we will see that 1000 likelihood evaluations are in most cases more than sufficient to recover the bias found in PE for MBHBs with $M\lesssim10^7\,M_\odot$, with sometimes as few as several hundred being sufficient (in which case, the optimization can take $\sim\!5$ seconds on a single core).

\subsubsection{Reparametrization}

We find that in instances where the optimizer can get stuck around local maxima, it is also useful to reparametrize the problem. This generally speeds up convergence and improves the robustness of our maximization procedure. Similarly to Ref.~\cite{Roulet:2022kot}, for the intrinsic parameters, we use a basis inspired by the waveform fitting literature. In particular, we work in terms of the logarithm of the mass ratio, the coefficient entering at the leading 1.5 order in the PN expansion for the spin $\psi_\chi$ (defined as in Eq.~(A2) of Ref.~\cite{Ng:2018neg}), and the ``antisymmetric spin'' $\chi_a$ (see, e.g., Refs.~\cite{Varma:2018aht,Varma:2018mmi}). The combination $\psi_\chi$ has been found in Ref.~\cite{Ng:2018neg} to be the best constrained spin parameter during the inspiral. %
For the extrinsic parameters, we use the cosine of the inclination angle and a combination similar to the ``chirp distance''~\cite{Fairhurst:2007qj} in place of the standard luminosity distance (we only include the dependence on the inclination and not the full detector pattern functions, which is equivalent to setting them to 1 in the definition in Ref.~\cite{Fairhurst:2007qj}, as we are mainly interested in mitigating the degeneracy between $D_L$ and $\iota$). This removes part of the degeneracy between distance and inclination for the quadrupole mode of the signal, which is generally (though, as we have shown, not always) dominant. Following Ref.~\cite{Roulet:2022kot}, less degenerate and multimodal variables could be defined also for the other extrinsic parameters, but given the different response of the LISA mission compared to ground-based interferometers, their definition is not straightforward. We defer this aspect to future work. To summarize, the variables we employ are: 
\begin{eqnarray}\label{eq:parameters_rep}
    \mathcal{M}_c&\rightarrow&\mathcal{M}_c\,,\nonumber\\ q&\rightarrow&\tilde{q}=\log q\,,\nonumber\\ 
    \chi_+&\rightarrow& \psi_\chi =  \dfrac{(92\chi_+-48\pi)(1+q) + 19\chi_-(q-1)}{128 q^{3/5}(1+q)^{-1/5}}\,,\nonumber\\
    \chi_-&\rightarrow&\chi_a=\frac{\chi_1-\chi_2}{2}\,,\nonumber\\
    D_L&\rightarrow&\tilde{D}_L=\frac{D_L}{\mathcal{M}_c^{5/6}}\left[\left(\frac{1+\cos^2\iota}{2}\right)^2+\cos^2\iota\right]^{-1/2}\,,\\
    \delta t&\rightarrow& \delta t\,,\nonumber\\
    \beta_L&\rightarrow&\beta_L\,,\nonumber\\
    \lambda_L&\rightarrow& \lambda_L\,,\nonumber\\
    \phi&\rightarrow&\phi\,,\nonumber\\
    \iota&\rightarrow&\cos \iota\,,\nonumber\\
    \Psi_L&\rightarrow&\Psi_L\,,\nonumber
\end{eqnarray}
where $\chi_\pm=\left(q\,\chi_1 \pm \chi_2\right)/\left(1+q\right)$. 
Note again that aside from the luminosity distance and inclination, we do not perform any transformation on the extrinsic parameters. The chirp mass is also left unchanged, as it is already a better-constrained parameter compared to, e.g., the total or individual masses, particularly for longer signals. In our optimization procedures, we rescale the parameters by their standard deviations obtained from the covariance matrix, ${\cal C}=\Gamma^{-1}$, which is calculated from the Fisher information matrix,
\begin{eqnarray}\label{eq:fisher_def}
    \Gamma_{ij}=\left(\frac{\partial h}{\partial \theta^i} \bigg|\frac{\partial h}{\partial \theta^j} \right)\,,
\end{eqnarray}
for the 11 original parameters,  $\theta^i=[\mathcal{M}_c, q,\chi_+,\chi_-, D_L,$ $\delta t,\beta_L,\lambda_L,\phi,\iota,\Psi_L]$.
When reparametrizing, the transformed covariance matrix, ${\cal C}'$, is given by ${\cal C}'=J\,{\cal C}\,J^{\rm T}$, with $J$ the Jacobian matrix associated to the transformation. %

\subsubsection{Using Fisher-informed priors}\label{sec:fisher_priors}

We also find that the optimization can be sped up significantly if we restrict the parameter space based on the approximate width of the posteriors determined via Fisher analysis. To accommodate the fact that there can be very severe bias due to waveform inaccuracy, we still generally allow for exploration up to $300\sigma_\theta$ in either direction of the injected value of each parameter, $\theta_{\rm inj}$. The resulting allowed parameter ranges are generally still considerably narrower than what was previously used in~\citetalias{Yi:2025pxe}. Of course, in cases where there is a physical limit on a given parameter's value which results in a narrower bound than $300\sigma_\theta$, we use the physical limit in the prior. For example, for the transformed mass ratio $\tilde{q}=\log q$ with $q$ bounded by $q=m_1/m_2>1$, the prior is taken to be 
\begin{equation}
[{\rm max}\left(0,\;\tilde{q}_{\rm inj}-300\sigma_{\tilde{q}}\right), \;\tilde{q}_{\rm inj}+300\sigma_{\tilde{q}}]\;.
\end{equation}

\begin{figure*}[htbp]
\centering
\includegraphics[width=0.99\textwidth]{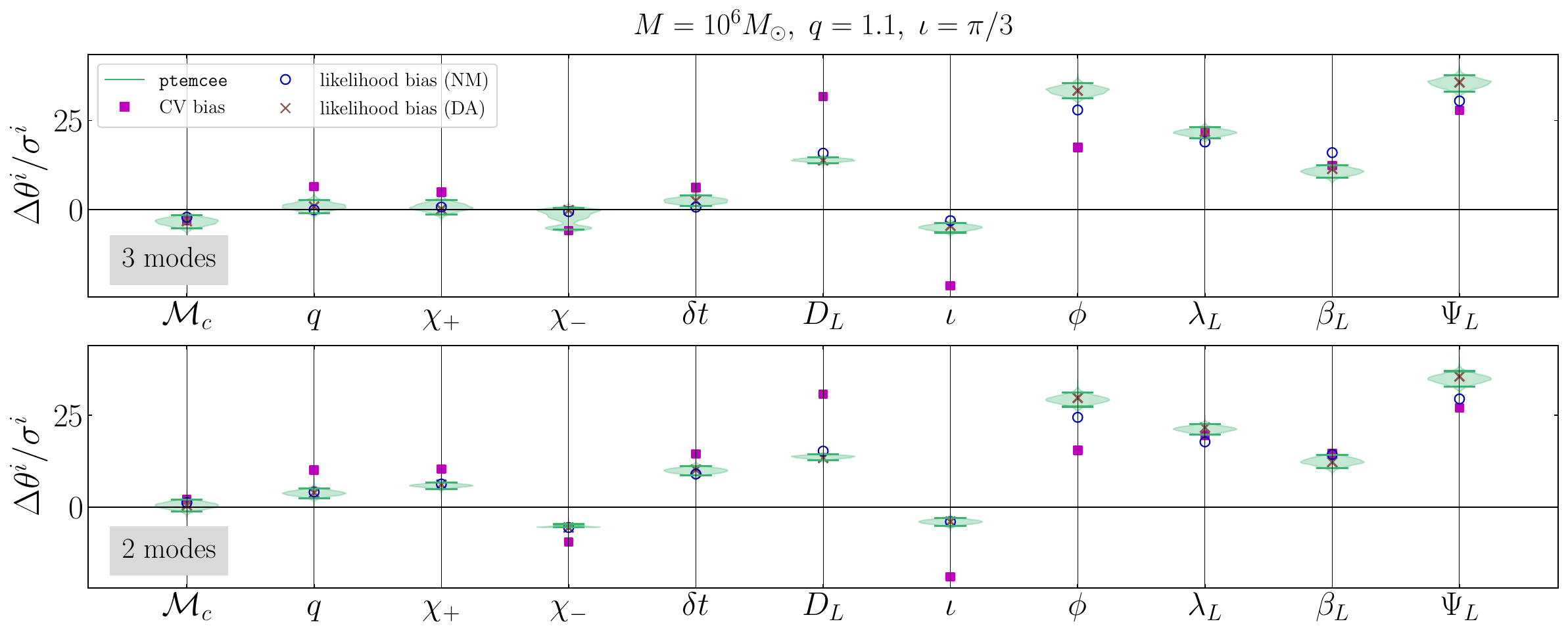} 
\caption{Same as Fig.~\ref{fig:cv_nm_1e7}, but comparing the simpler likelihood optimization method (``NM'' for Nelder-Mead algorithm only, as in~\citetalias{Yi:2025pxe}) with the improved likelihood optimization introduced in Sec.~\ref{sec:dual_annealing} (``DA'' for dual annealing). Aside from being more robust, the latter method also occasionally performs better in recovering the bias found in PE for some parameters, particularly in cases of more severe biases. In the top panel, signal recovery is performed with 3 modes $=\{(2,\,2),\,(3,\,3),\,(2,\,1)\}$ (excluding the $(4,\,4)$ and $(3,\,2)$), and in the bottom panel, the template contains only $\{(2,\,2),\,(3,\,3)\}$ (neglecting $\{(2,\,1),\,(4,\,4),\,(3,\,2)\}$). }
\label{fig:nm_vs_annealed}
\end{figure*}

With these modifications to the original optimization procedure using a basic implementation of the Nelder-Mead algorithm, we find that we are generally able to recover the correct systematic biases due to waveform inaccuracies, even in the limit of much louder signals and more extreme biases. 
We demonstrate this for a couple of cases in Fig.~\ref{fig:nm_vs_annealed}, where we show significant biases due to neglecting 2 or 3 higher-order modes with respect to the true template. In these cases, the simple Nelder-Mead method (``NM'', blue circles) employed in~\citetalias{Yi:2025pxe} struggles to correctly estimate the bias on the extrinsic parameters, whereas the new likelihood optimization method with dual annealing, reparametrization, etc. (``DA'', brown crosses) performs better. In general, apart from the fact that the dual annealing is a global optimization method and is therefore designed to be more effective over multimodal surfaces, we also take advantage of the fact that it generally makes our likelihood optimization method more robust, albeit at a slightly higher computational cost. By noting when multiple initializations do not return the same maximum value, we have a sense of when a likelihood is particularly difficult to optimize, and when we should therefore be careful in taking the results to be trustworthy. 

Finally, we note that modest improvement is achieved by using regular Nelder-Mead with just the reparametrization and addition of Fisher-informed priors, i.e., without the dual annealing, even in the regime of high-mass MBHBs, which we find to be the most delicate. In Sec.~\ref{sec:utility_of_dual_annealing}, we will see that using all of the above improvements allows us to find the true maximum log-likelihood in a given region of parameter space at least $\sim\!40$\% of the time in the regime of high-mass MBHBs (see Fig.~\ref{fig:lnL_by_octant}). Using just the reparametrization and restricted priors allows us to find the correct maximum about $\sim\!10-30$\% of the time, albeit with considerably fewer likelihood evaluations (as low as 100-200, and nearly always less than 1000). For the remainder of this work, we use all improvements to the original Nelder-Mead optimization mentioned in this section to obtain as robust of a result as possible, but we note here that depending on the problem at hand, one could choose to leave out the dual annealing to obtain faster results that can still be fairly robust with several initializations. Using just the original Nelder-Mead approach, we are often unable to find the true maximum even once in five initializations.

\section{Spin dependence of systematic biases}\label{sec:spin_dependence}

In~\citetalias{Yi:2025pxe}, the analysis was restricted to MBHBs with zero progenitor spins. In this section, we use the improved likelihood optimization techniques explored in the previous section to investigate the severity of systematic bias due to neglected higher-order modes as a function of spin. We expect significant effects given the spin corrections entering at next-to-leading-order (NLO) in the mode amplitudes, especially for $\ell\neq m$ modes (see, e.g., Refs.~\cite{Berti:2007nw,Mishra:2016whh,Garcia-Quiros:2020qpx}). 

\begin{figure*}[htbp]
\centering
\begin{tabular}{l@{\hskip .4cm}l}
    \includegraphics[width=0.45\textwidth]{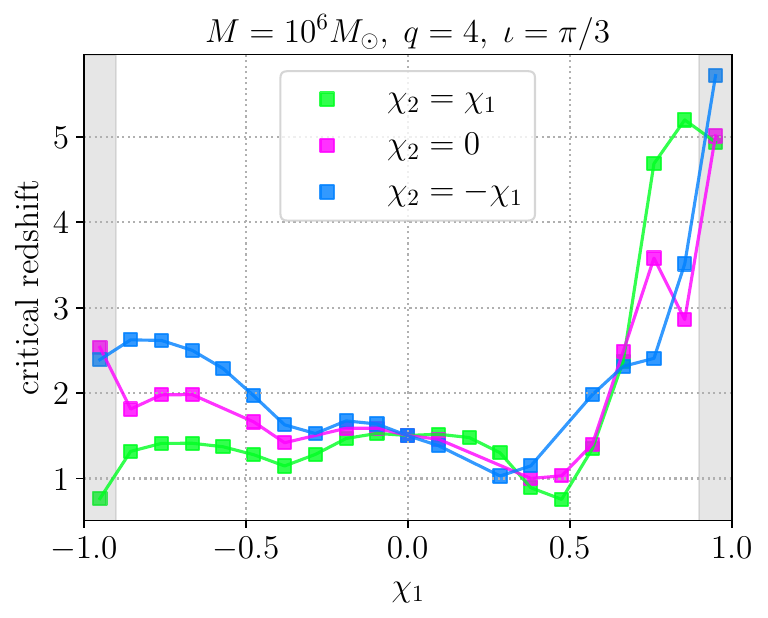} & \includegraphics[width=0.45\textwidth]{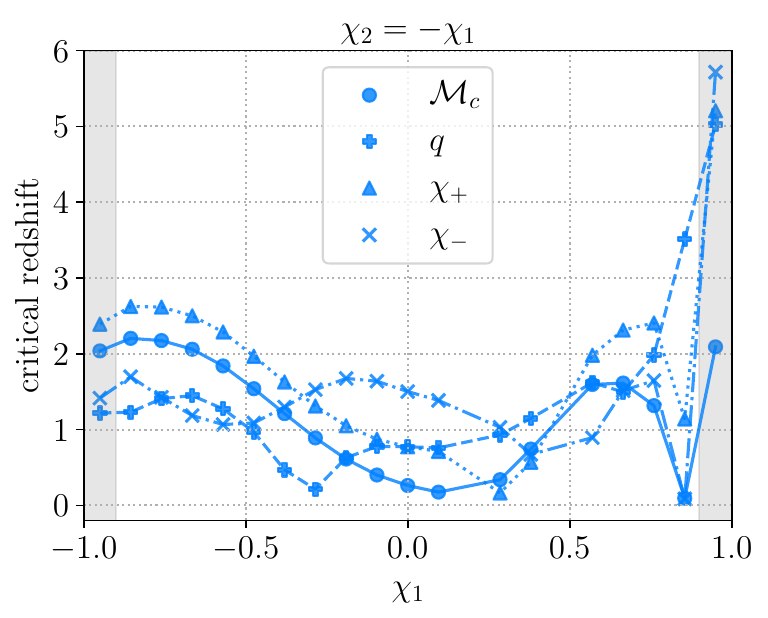} \\
    \includegraphics[width=0.45\textwidth]{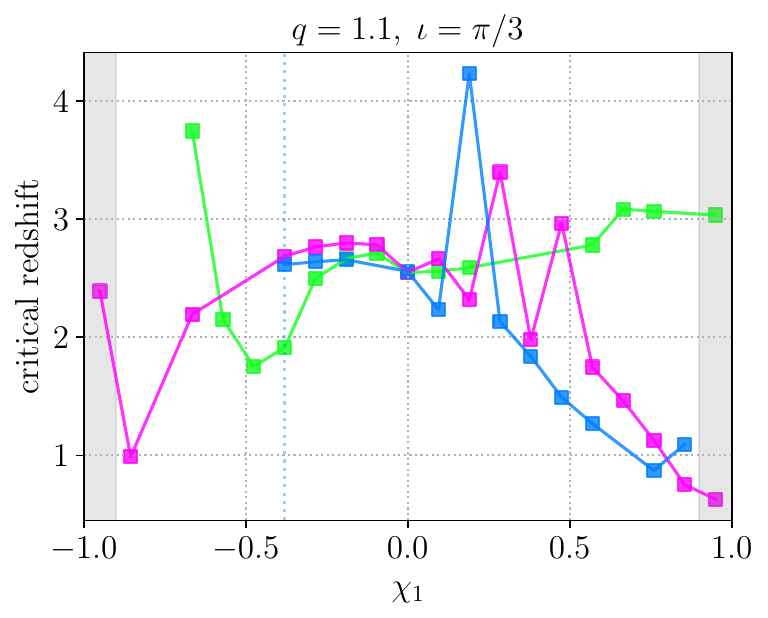}  & \includegraphics[width=0.45\textwidth]{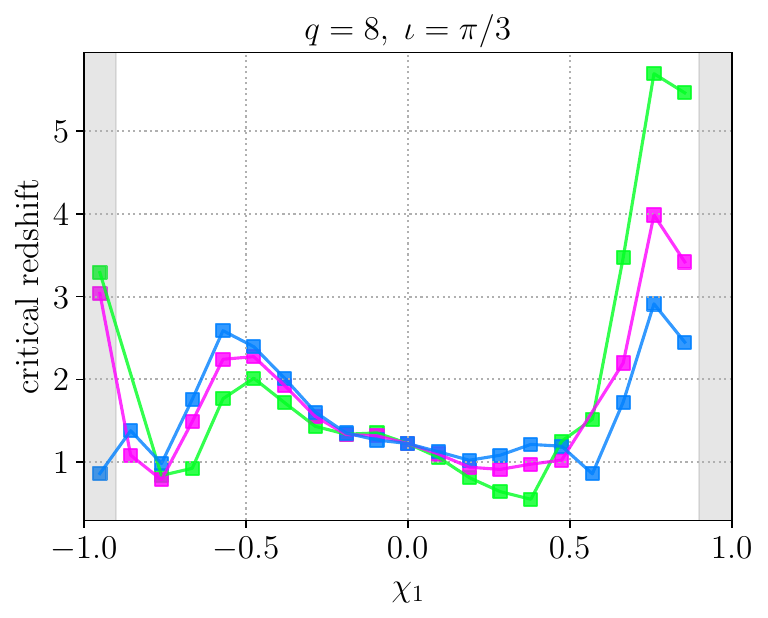} \\
    \includegraphics[width=0.45\textwidth]{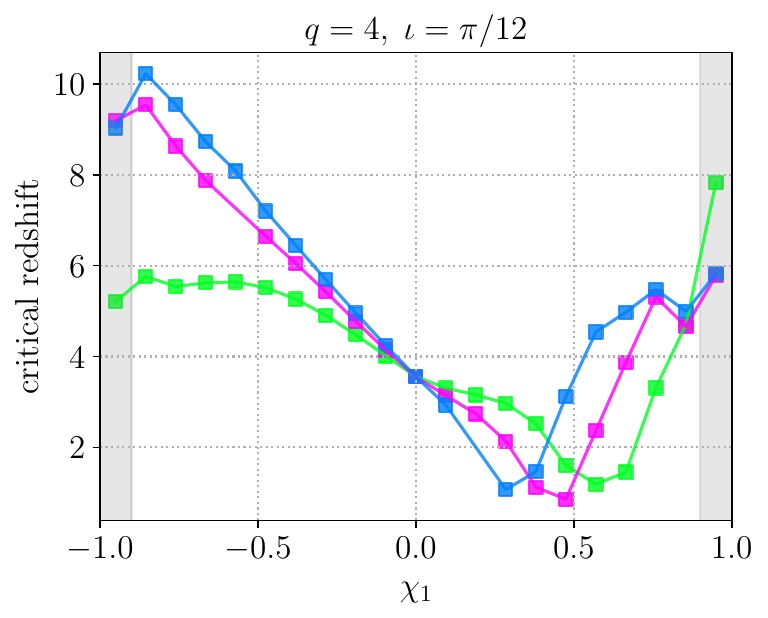} & \includegraphics[width=0.45\textwidth]{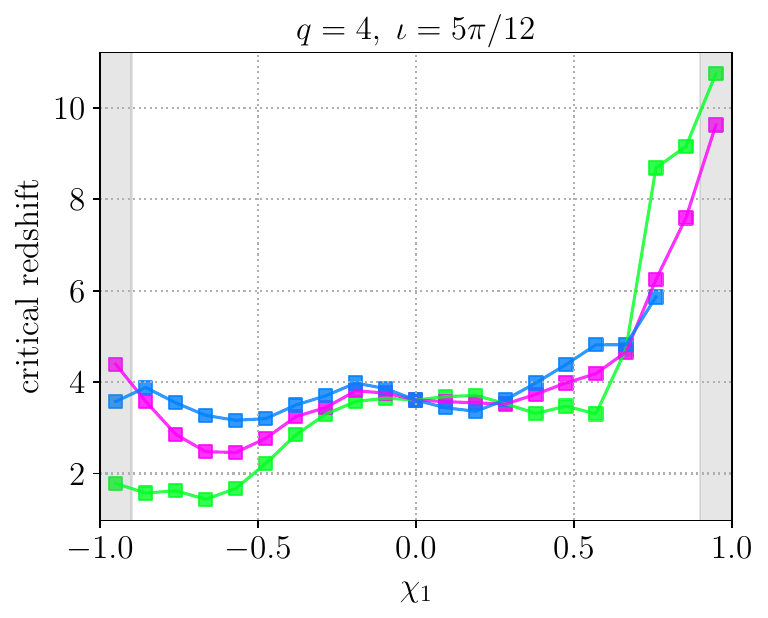} 
\end{tabular}
\caption{Critical redshift at which the bias on intrinsic parameters due to neglecting the (3,\,2) mode becomes greater than the $2\sigma$ statistical error. We show results for the 
(3,\,2) mode as a representative case, since it is among the weakest higher-order modes; the resulting bias therefore provides a conservative illustration of the impact of neglecting subdominant modes. We plot the minimum redshift as a function of $\chi_1$, with different colors corresponding to different values of $\chi_2$. In the top left panel, we show the critical redshift for a reference MBHB with $M=10^6\,M_\odot,\,q=4,\,\iota=\pi/3$. The top right panel breaks down the critical redshift for unbiased inference of individual intrinsic parameters for the $\chi_2=-\chi_1$ case (blue) shown in the top left panel. In the second and third rows, we show results with different mass ratios and inclinations with respect to the reference MBHB, respectively. %
}
\label{fig:bias_vs_spin}
\end{figure*}

In Fig.~\ref{fig:bias_vs_spin}, we plot the critical redshift at which the bias on intrinsic parameters $(\mathcal{M}_c,\,q,\,\chi_+,\chi_-)$ due to neglecting the (3,\,2) mode becomes larger than the $2\sigma$ statistical error on the same parameters. This redshift is plotted as a function of $\chi_1$, with $\chi_2$ taken to be either equal to $\chi_1$, zero, or equal in magnitude but opposite in sign (i.e., antiparallel) to $\chi_1$. Above the critical redshift lines, the SNR is low enough that the statistical error on the parameters dominates. In the top left panel, we show how this critical redshift depends on spin for a set of reference MBHB parameters ($M=10^6M_\odot,\,q=4,\,\iota=\pi/3$). The top right panel shows the critical redshift for significant bias on each individual intrinsic parameter for the same reference MBHB parameters. For ease of visualization, we focus here on only $\chi_2=-\chi_1$ systems (marked in blue in the left plot), and each linestyle/plot marker style corresponds to the critical redshift for a different parameter. Here, we see that depending on the spin of the binary, different parameters can be more or less biased than others given the same total mass, mass ratio, and extrinsic parameters (in this case, the bias on either $\chi_+$ or $\chi_-$ tends to dominate). We compute statistical errors from the Fisher matrices and use the improved likelihood optimization method to compute systematic errors. 

In the second row of Fig.~\ref{fig:bias_vs_spin}, we show the critical redshift as a function of spin when changing the mass ratio with respect to the top left panel ($q=1.1$ and 8 on the left and right, respectively). Finally, in the third row, we change the inclination angle with respect to the top left panel. We note that although the \textsc{IMRPhenomXHM} waveform was calibrated to high
spins, relatively few numerical relativity waveforms were available in this
regime during its calibration, particularly for the first version (``version \texttt{122019}''), which is used in \texttt{lisabeta}. We are therefore less confident in the results
shown for $|\chi_{1,2}|\gtrsim0.9$, and have grayed out these regions in the
plots.

To make the optimization more robust against finding incorrect maxima, for each data point in Fig.~\ref{fig:bias_vs_spin} we initialize the optimizer several times, each time drawing an initial guess from the multivariate normal distribution centered at the injected parameters and with covariance given by the inverse Fisher information matrix. If the same maximum is not found between several different iterations, we assume the maximum cannot be reliably located by the algorithm and do not plot a point. By finding the ``same maximum,'' we mean that we demand that each of the 11 parameters exhibit no more than a 10\% difference as evaluated at the different maximum likelihood values. %
We have also verified that repeating the maximization for different initializations generally allows us to find a maximum close to the one found by full PE within a handful of iterations. We thus consider a maximization converged if the same maximum is found for at least 2 different iterations among 5. In one particular region, i.e. for low $\chi_1$ in the blue $\chi_2=-\chi_1$ case in the $q=1.1, \, \iota=\pi/3$ panel, the optimizer struggled significantly. Upon examining the waveforms in this region, we found unexpected behavior in the $(3,\,3)$ mode amplitude (see Appendix~\ref{app:compare_wfs}). We have therefore excluded points to the left of $\chi_1=-0.38$ for this line (marked by a dotted blue vertical line).  %

In general, it is clear that individual spins have a significant effect on the extent to which biases will be observed due to neglecting the (3,\,2) mode in various MBHB systems; there is non-negligible change going from left to right in each panel. Notably, for $q\geq4$, the trends are qualitatively similar between the green, pink, and blue lines, i.e., no matter which way $\chi_2$ is oriented with respect to $\chi_1$. This is consistent with our findings in, e.g., Fig.~\ref{fig:spin_dominance_condensed}, where we see the relative contribution of each mode to depend mostly on $\chi_1$ for large enough mass ratios. In general, the plots in Fig.~\ref{fig:bias_vs_spin} highlight the fact that the relative bias compared to statistical error cannot be trivially predicted, but rather depends sensitively on many factors, including the mass ratio, inclination, and shape of the noise curve. 
We also see the intricacy of different higher-order modes' contributions to the total GW signal: each of the modes' contributions to the overall signal can increase or decrease 
at different values of the spins compared to the other modes, leading to nonmonotonic behavior of the critical redshift in many places.

There are particularly noticeable ``jumps'' in the critical redshift between
$\chi_1 \sim 0.1$ and $\chi_1 \sim 0.5$ in the $q=1.1 \, ,\iota=\pi/3$ case. We first rule out
failures of the likelihood-optimization procedure by validating these points
with full PE runs; indeed, if anything, the Fisher matrices can underestimate the statistical error on $\chi_-$, such that the spikes are even a bit more pronounced when we make the same plot from the PE samples. Upon looking at the waveforms in this regime, we then find that these jumps coincide with pronounced dips in the
frequency-domain amplitude of the $(2,\,1)$ mode at the same spin values. We
attribute this behavior---which we observe across several waveform approximants,
including \textsc{IMRPhenomHM}~\cite{London:2017bcn} and
\textsc{SEOBNRv5HM\_ROM}~\cite{Pompili:2023tna}---to the PN structure of the
$(2,\,1)$ amplitude. Specifically, the LO contribution to the $(2,\,1)$
mode enters at $0.5$\,PN order relative to the leading $(2,\,2)$ term and is
proportional to the mass difference, while the NLO
contribution enters at $1$\,PN order and depends on the spins.
For nearly equal-mass systems, the coefficient of this NLO
term is dominated by the antisymmetric spin combination and may have the
opposite sign to the LO term, which is itself suppressed in this
limit. This gives rise to a competition between the LO and NLO contributions during the inspiral, producing dips in the
$(2,\,1)$ amplitude as a function of frequency (see Appendix~\ref{app:compare_wfs} for plots of this behavior as modeled by different waveform approximants).
In the equal-mass limit, the spin-dependent term becomes the LO PN
contribution to the $(2,\,1)$ mode and depends solely on the antisymmetric spin
combination, which is not suppressed by the mass difference.
Consequently, even for nearly equal masses, the $(2,\,1)$ mode can contribute
appreciably relative to the $(2,\,2)$ mode, as it is the dominant multipole
associated with spin effects. This behavior is consistent with the PN order
counting discussed in Sec.~\ref{sec:spin_mode_ordering} and in (e.g.)
Ref.~\cite{Berti:2007nw}.
As a result, variations in the spins can lead to substantial changes in the
$(2,\,1)$ contribution, thereby significantly affecting the magnitude of the
resulting systematic bias. In particular, for the $q=1.1,\,\iota=\pi/3$ case shown in Fig.~\ref{fig:bias_vs_spin}, the amount of bias due to removing the $(3,\,2)$ mode depends significantly on the aforementioned features of the $(2,\,1)$ mode at different values of the spins. Notably, we have checked that the ``spikes'' in, e.g., the $\chi_2=0$ case disappear if we inject without the $(2,\,1)$ mode and recover without either the $(2,\,1)$ or the $(3,\,2)$, which further highlights the impact of the $(2,\,1)$ mode here.  

In summary, we find in Fig.~\ref{fig:bias_vs_spin} that the effect of varied progenitor spins on the degree of systematic bias due to a neglected higher mode is both significant and nontrivial. This will be an important finding to consider in ongoing studies of which additional, currently unmodeled higher modes will ultimately be required for unbiased analysis of MBHBs observed by LISA. Our findings suggest that in trying to predict biases due to $\ell \geq6$ and various $\ell \neq m$ modes, it will not be sufficient to consider only the effects of total mass, mass ratio, and inclination. Rather, one must also account for the impact of progenitor spins in conjunction with these other effects. Neglecting the dependence on $\chi_{1,2}$ can cause one to significantly underestimate the bias due to a neglected mode at large spin magnitudes (see, e.g., the bias at $\chi_1<-0.7$ and $\chi_1>0.7$ compared to $\chi_1 =0$ in the bottom left and bottom right panels of Fig.~\ref{fig:bias_vs_spin}, respectively).

Finally, it is important to note that the overall accuracy of current waveform approximants may not be sufficient for analyzing high-SNR events observed with LISA. The mismatches between \textsc{IMRPhenomXHM} and NR surrogate waveforms with higher modes are around $10^{-3.5}-10^{-1.5}$ with $q<9.09$ and spin magnitudes $<0.8$, and somewhat higher outside this range of parameters (see Figs.~14 and 15 of Ref.~\cite{Garcia-Quiros:2020qpx}). Moreover, it has been shown that incomplete higher-order PN terms in current phenomenological waveforms can cause significant systematic errors for high-SNR events observed with even current ground-based detectors~\cite{Owen:2023mid}. Indeed, even current NR waveforms may not be sufficiently accurate for the much larger SNRs expected for next-generation observations; in Ref.~\cite{Jan:2023raq}, errors due to finite grid resolution were demonstrated to be limiting for events with $q\gtrsim6$ as seen by Cosmic Explorer~\cite{Reitze:2019iox}. There is much ongoing work in understanding the accuracy requirements for LISA of not only waveform approximants (such as the \textsc{IMRPhenomXHM} model used here), but also the NR waveforms to which these approximants are themselves calibrated (see a discussion of the readiness of the latest SXS catalog for next-generation detectors in Ref.~\cite{Scheel:2025jct}). That being said, our goal here is to quantify the relative effect of higher-order modes. Provided that the \textsc{IMRPhenomXHM} model used here does not substantially
overestimate or underestimate the contribution of higher-order modes to the
total waveform, we expect our qualitative conclusions to remain valid even as
waveform models continue to improve in accuracy over the coming decades.

\section{Finding degenerate positions in the sky}\label{sec:degenerate_sky_pos}

One of the most interesting aspects of systematic biases in the very high-mass, high-SNR regime is the fact that these biases can change the hierarchy of log-likelihood values between nearly degenerate positions in the sky. These degeneracies are exact in certain approximations of the LISA response~\cite{Baibhav:2020tma}. Their impact, and the way in which the full LISA response can break them, is discussed in Ref.~\cite{Marsat:2020rtl} (see also Ref.~\cite{Mangiagli:2022niy}). In total, there are eight nearly degenerate positions, which can be parametrized by indices $(s,n)$ describing the transformations 
\begin{eqnarray}
\label{eqn:octants}
    \beta_L^{(s)} &\rightarrow& s\times\beta_L\,,\\ 
    \lambda_L^{(n)} &\rightarrow&\lambda_L + n\times\pi/2\,,\nonumber
\end{eqnarray}
where $s$ takes the value of either $-1$ or $1$, and $n\in[0,1,2,3]$. We will often refer to a particular ``octant'' in the sky, meaning a pair of $(s,n)$ in relation to a true sky position. 

\begin{figure*}[htbp]
\centering
\includegraphics[width=0.96\textwidth]{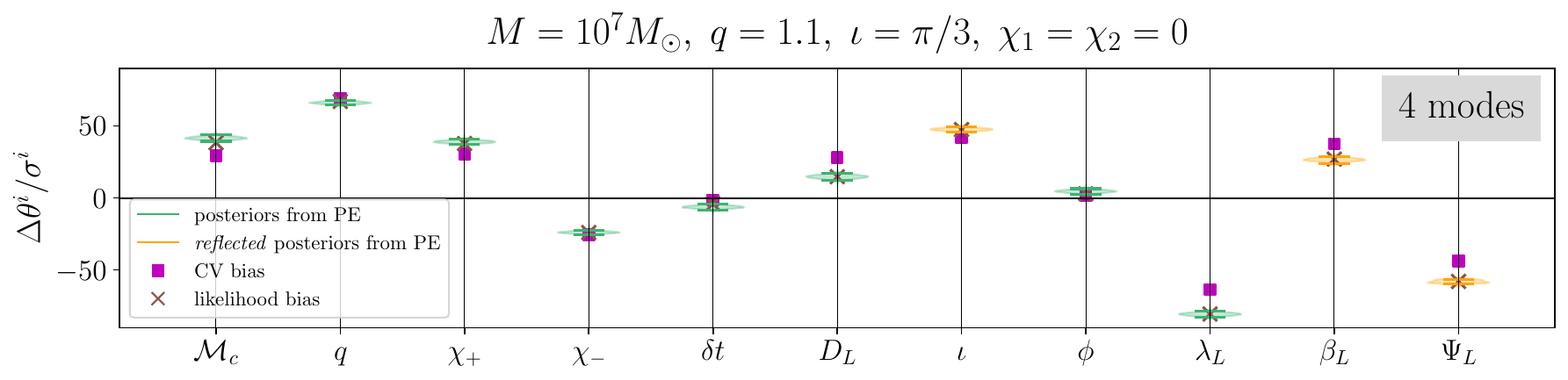} 
\caption{Same as Fig.~\ref{fig:cv_nm_1e7} for an MBHB of the same total mass as those in Fig.~\ref{fig:cv_nm_1e7}, but with a more asymmetric mass ratio, and therefore shorter signal duration. In this case, the PE with a higher-order mode excluded finds the \emph{reflected} sky position mentioned in Ref.~\cite{Marsat:2020rtl}: see Eq.~(63) of this reference, as well as Fig.~\ref{fig:lnL_by_octant_skymap} below. The direct likelihood optimization method finds a bias from the injected value that is equal in magnitude to the bias between the PE posteriors and the true \emph{reflected} position. To illustrate this, we plot the reflected values of the posteriors on $\iota\;,\,\beta_L,$ and $\Psi_L$ in orange violin plots. The actual posteriors found by the \texttt{ptemcee} sampler are biased from the reflected positions of injected values as defined in Eq.~\eqref{eq:reflected_mode} (i.e., well outside the range of the plots).}
\label{fig:cv_nm_1e7_reflected_mode}
\end{figure*}

In~\citetalias{Yi:2025pxe}, the masses considered were low enough that the motion of the LISA detector and/or its high-frequency response generally broke the near-degeneracies between the different octants, leaving only a single posterior on $(\lambda_L,\beta_L)$ in the same octant as the injected signal. The more massive binaries considered here, however, have shorter signals and merge at lower frequencies: 
$f_{\rm ISCO}=2.2\times10^{-3}\,{\rm Hz}$ and $2.2\times10^{-5}\,{\rm Hz}$ for $10^6\,M_\odot$ and $10^8\,M_\odot$ systems, respectively, for the frequency of the (2,\,2). As a result, neither the motion of LISA's detector nor its high-frequency response are as effective at breaking the degeneracies in the sky position, leaving multimodalities in $(\lambda_L,\beta_L)$. In some cases, the PE can converge on the \emph{reflected} mode discussed in Ref.~\cite{Marsat:2020rtl}, where the extrinsic parameters are transformed as 
\begin{eqnarray}\label{eq:reflected_mode}
    \lambda_L^* &\rightarrow&\lambda_L\,,\nonumber\\
    \beta_L^* &\rightarrow&-\beta_L\,,\nonumber\\
    \Psi_L^* &\rightarrow& \pi-\Psi_L\,,\\
    \iota^* &\rightarrow& \pi - \iota\,,\nonumber\\
    \phi^* &\rightarrow& \phi \,.\nonumber
\end{eqnarray}
In the ``frozen'' LISA approximation~\cite{Marsat:2018oam}, i.e., where one takes the detector to be in a constant position throughout the observation of the event, the degeneracy between the true and \emph{reflected} positions is exact~\cite{Baibhav:2020tma,Marsat:2020rtl}, causing the \emph{reflected} mode to appear in a significant number of MBHB events. (Relativistic effects can actually break this exact degeneracy, although this is only expected to be relevant for the loudest signals~\cite{vanderSteen:2025nxy}.) For yet more massive binaries or shorter signals (chirp mass $\mathcal{M}_c\gtrsim10^7\,M_\odot$; see Fig.~13 of Ref.~\cite{Mangiagli:2022niy}), one can often find the additional six spurious modes as well.

This leads us to an area in which systematic biases can be very dangerous: for the short, low-frequency signals of very massive MBHBs, neglecting a higher-order mode can introduce enough of a difference in the waveform as to result in the inference of an entirely incorrect position in the sky. This phenomenon is illustrated in Fig.~\ref{fig:cv_nm_1e7_reflected_mode}, where, similar to Fig.~\ref{fig:cv_nm_1e7}, we show the Cutler-Vallisneri and direct likelihood optimization estimates of systematic bias compared to the actual biases recovered in PE for an event with $M=10^7\,M_\odot$. 
For the parameters $\iota,\;\beta_L,$ and $\Psi_L$, we plot in orange the \emph{reflected} posteriors on $\beta^*_L \rightarrow -\beta_L,\;\iota^* \rightarrow \pi - \iota,\;\Psi_L^* \rightarrow \pi-\Psi_L$, where quantities without asterisks correspond to the values found in PE. That is, the maximum likelihood found by the PE is in an entirely different region of the sky, at the \emph{reflected} mode, which is well outside of the range of the plots shown. The ``reflection'' of that \emph{reflected} mode, shown in orange here, lands us back in the same octant containing the injected parameters. For the event shown in Fig.~\ref{fig:cv_nm_1e7_reflected_mode}, there are also very small secondary peaks in the original octant (nonreflected positions). However, these consist of only a few samples (87 out of $3.8\times10^4$), demonstrating that a neglected higher-order mode can lead to confident inference of an incorrect sky position posterior. 
Note that in this instance, the likelihood optimization algorithm finds the \emph{secondary} maximum, i.e., the one in the same octant that contains the injected parameters. 
Notably, the Cutler-Vallisneri estimates are also moderately successful in generally finding at least the correct sign of the systematic biases. 

\begin{figure*}[htbp]
\centering 
\includegraphics[width=0.48\textwidth]{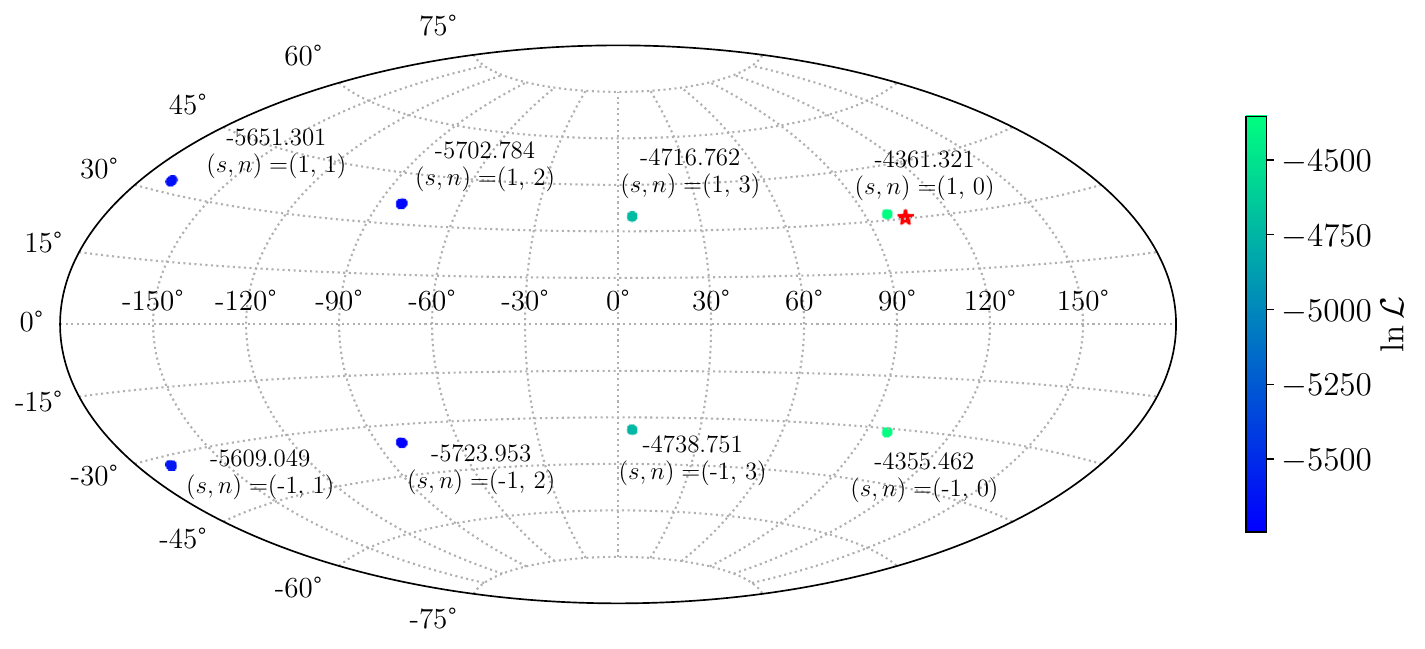}
\includegraphics[width=0.48\textwidth]{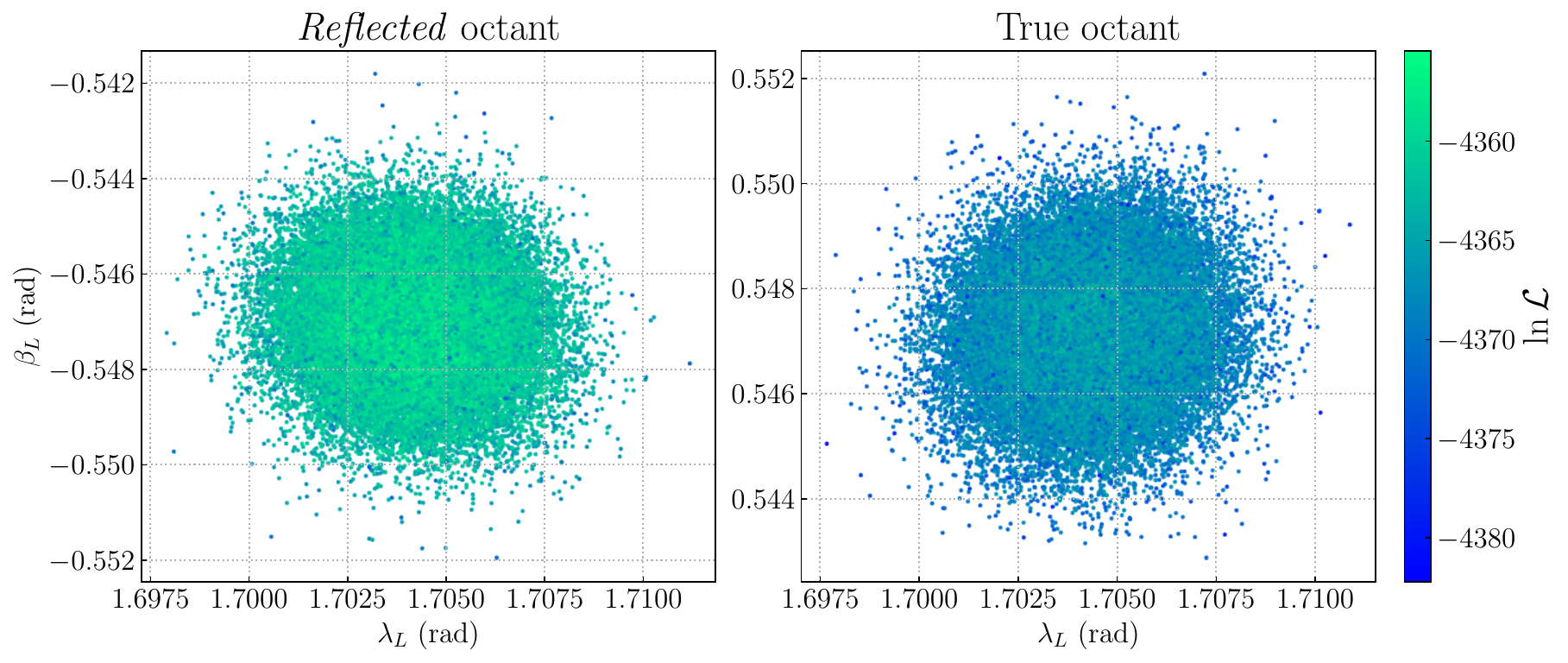}
\caption{The waveform error due to a neglected higher-order mode ((3,\,2) in this case) can cause mislocalization of an MBHB. Left: Posterior samples on the sky position parameters $(\lambda_L,\beta_L)$ as recovered in PE restricted to each octant in the sky as defined in Eq.~\eqref{eqn:octants}. In each octant, we report the local maximum $\ln \mathcal{L}$ value found in PE. The samples are colored by log-likelihood values, and the injected sky position is marked by a red star. We see that the log-likelihood is highest in the true octant with $(s,n) = (1,\,0)$ and the \emph{reflected} octant with $(s,n) = (-1,\,0)$. Right: Focusing on the two octants with the highest local $\ln \mathcal{L}_{\rm max}$, we see that the log-likelihood is actually slightly higher in the \emph{reflected} sky position. %
}
\label{fig:lnL_by_octant_skymap}
\end{figure*} 

We stress here that the convergence of the \texttt{ptemcee} sampler on the \emph{reflected} mode is not a result of the PE failing.
Rather, when excluding the $(3,\,2)$ mode, the recovery waveform model becomes
sufficiently different from the true waveform model that the maximum-likelihood point,
when projected onto the $\{\lambda_L,\,\beta_L,\,\Psi_L,\,\iota,\,\phi\}$ subspace,
coincides with a position near (i.e., slightly biased from) the \emph{reflected} sky position of the injected source.
To verify this, we run PE restricting the sampler to one octant in the sky at a time, with each octant specified by $(s,\,n)$ as defined in Eq.~\eqref{eqn:octants}. This allows us to compare the local maximum likelihood value found by PE in the true sky mode---i.e., $(s,\,n) = (1,\,0)$---with that found in one of the nearly degenerate modes. For instance, for the nonspinning $M=10^7\,M_\odot,\;q=1.1,\;\iota=\pi/3$ event shown in Fig.~\ref{fig:cv_nm_1e7_reflected_mode}, we find in different octants the maximum log-likelihood ($\ln \mathcal{L}$) values shown in Fig.~\ref{fig:lnL_by_octant_skymap}. On the left, we plot on a sky map the posterior points on $\lambda_L$ and $\beta_L$ found in the octant-restricted PE, with points colored by the log likelihood of the samples. A red star marks the injected sky position. Note how well the points are localized, given the mass and SNR of the system ($\rho>3000$). We see that the $\ln \mathcal{L}_{\rm max}$ found in the true and \emph{reflected} modes is quite similar, with the other modes having significantly lower values. 

On the right side of Fig.~\ref{fig:lnL_by_octant_skymap}, we focus on the posteriors in just the \emph{reflected} and true octants. Here, it is clear that the $\ln \mathcal{L}_{\rm max}$ found in the $(-1,\,0)$ octant is indeed slightly higher, with the difference between local maximum values being $\Delta \ln \mathcal{L}_{\rm max}=5.859$. 
Notably, while this is a smaller difference than we might anticipate from noise fluctuations (see Eq.~\eqref{eqn:logL_criterion}), it nevertheless results in significant changes to the reconstructed parameters. This highlights the fact that in this very high-SNR regime, one must be careful to mitigate even errors that seem small according to criteria such as Eq.~\eqref{eqn:logL_criterion}, if one wishes to accurately localize and characterize an MBHB signal with LISA. 

Finally, we note that even without restricting to a given octant in the sky, the \texttt{ptemcee} sampler found the \emph{reflected} sky mode with $\ln \mathcal{L}_{\rm max}=-4355.35$. This further confirms that the fact that PE has found the reflected mode is not a failure of the sampler; rather, removing the (3,\,2) mode alters the template significantly enough that the $\ln \mathcal{L}_{\rm max}$ is truly located in a different octant of the sky with respect to the one containing the injected parameters. We note that a similar phenomenon can happen with different kinds of systematic errors other than neglected higher-order modes, in particular inaccuracies in the waveform approximants themselves for a fixed physical content~\cite{Marsat_inprep}. %

\section{Direct likelihood optimization to address multimodalities}\label{sec:utility_of_dual_annealing}

In the previous section, we found that the difference in $\ln \mathcal{L}_{\rm max}$ values between sky octants can be small, particularly for shorter MBHB signals. This brings us to an additional manner in which the direct likelihood optimization method can be helpful in analyzing very massive events, where the hierarchy between different octants in the sky is not clear \emph{a priori}. We find that if we restrict the likelihood optimization algorithms to a given octant in the sky at a time, we can often recover not only the correct hierarchy between different octants, but also the respective $\ln \mathcal{L}_{\rm max}$ values recovered in PE. 
\begin{figure*}[htbp]
\centering 
\includegraphics[width=0.98\textwidth]{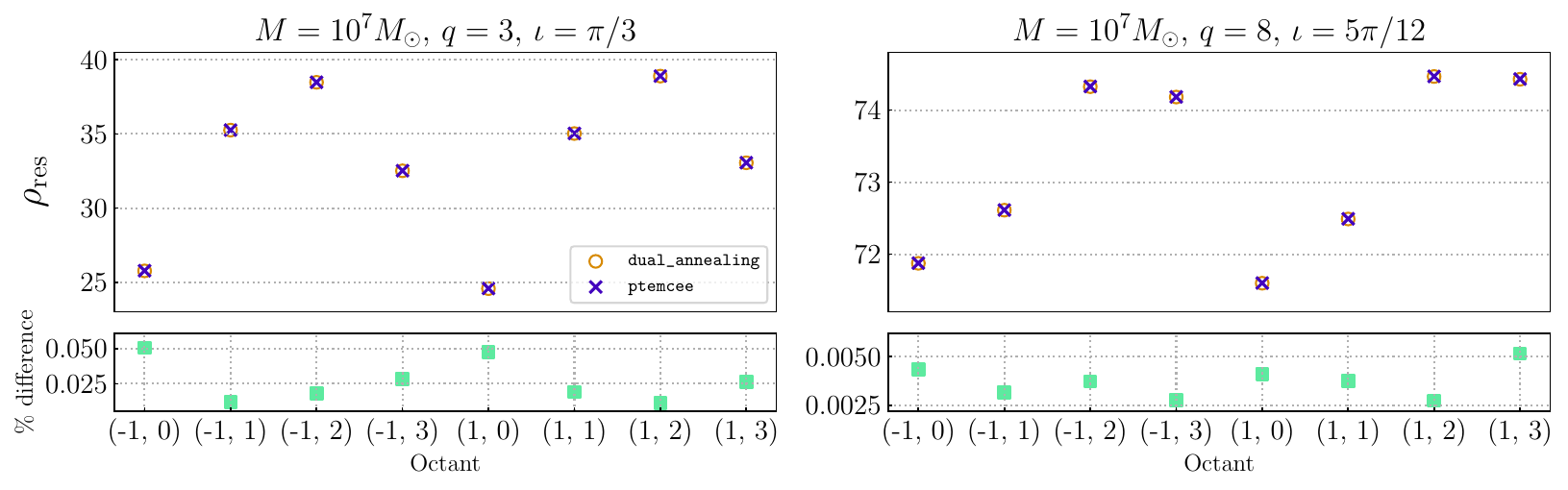} 
\caption{ Residual SNR $(\rho_{\rm res}=\sqrt{-\ln \mathcal{L}_{\rm max}})$ as found in different octants of the sky by the direct likelihood optimization method (orange circles) and PE (dark purple crosses). The values agree remarkably well, as can be seen in the percentage difference between $\rho_{\rm res}$ values plotted in the bottom panels. This suggests that the direct likelihood optimization method can be used to reduce the expense of running PE for very massive BBHs, where there are multimodalities in the sky angle posteriors. %
}
\label{fig:rho_res_dlo_vs_pe}
\end{figure*}
This can be seen in Fig.~\ref{fig:rho_res_dlo_vs_pe}, where for a few very massive events, we show the comparison of $\rho_{\rm res}=\sqrt{-\ln \mathcal{L}_{\rm max}}$~\cite{Cornish:2011ys,Chandramouli:2024vhw} in each octant as found by PE and the direct likelihood optimization method using all the refinements discussed in Sec.~\ref{sec:dual_annealing}. 
We note here that $\ln \mathcal{L}_{\rm max}$ corresponds to the local maximum in each octant, and therefore $\rho_{\rm res}$ corresponds to a local (in each octant) estimate of the residual SNR.

\begin{figure}[htbp]
\centering 
\includegraphics[width=0.48\textwidth]{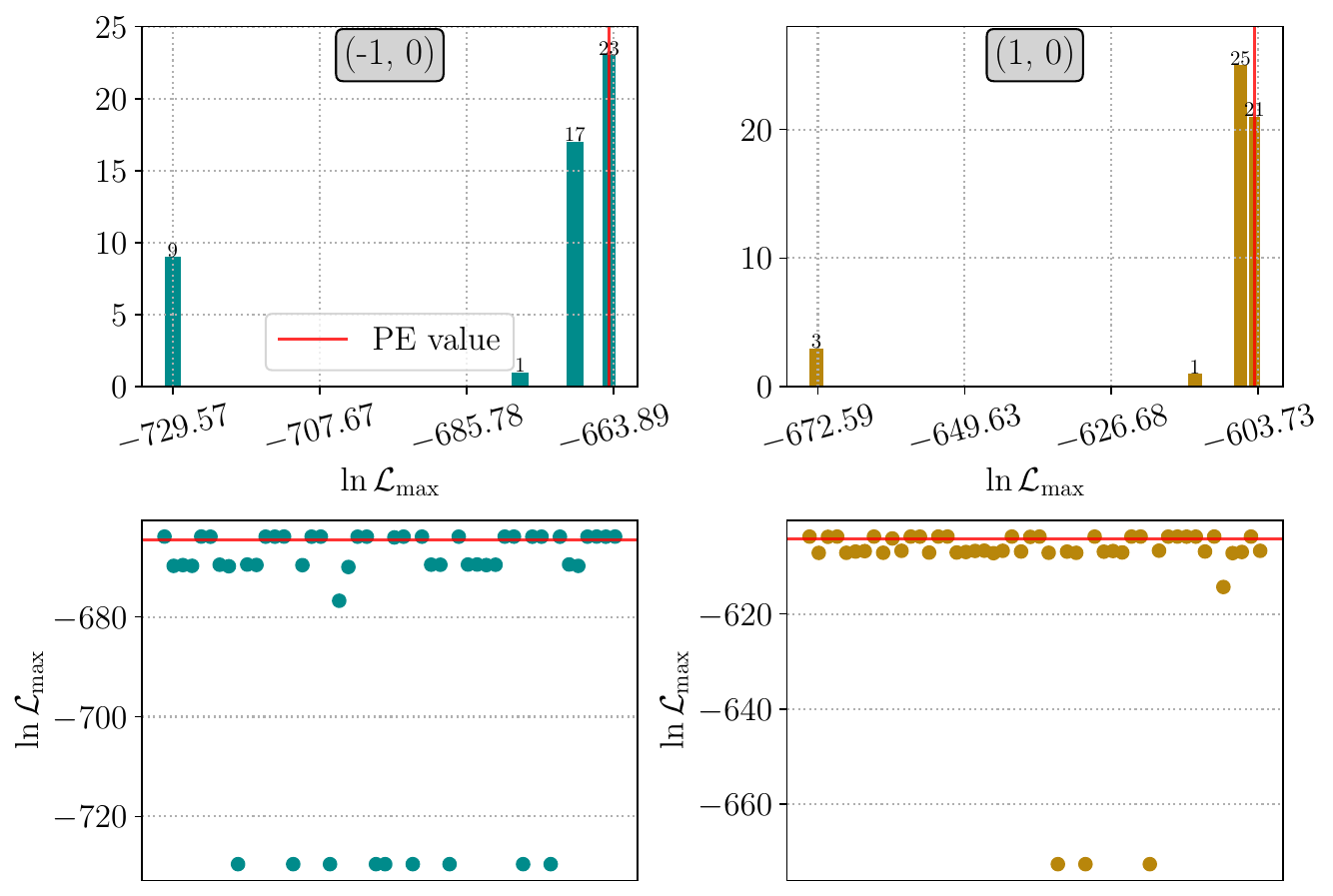} 
\caption{Histogram and scatter plot of the maximum log-likelihood value found by our direct likelihood optimization routine for an event with $M=10^7\,M_\odot,\;q=3,\;\iota=\pi/3$ over multiple iterations. The procedure generally finds the same $\ln \mathcal{L}_{\rm max}$ found in PE (i.e., the least negative value).}
\label{fig:lnL_by_octant}
\end{figure}

We initialize the optimizer several times in the same manner as in Sec.~\ref{sec:degenerate_sky_pos}. The values shown in  Fig.~\ref{fig:rho_res_dlo_vs_pe} are the largest values found after initializing five times. Due to the higher masses compared to Fig.~\ref{fig:bias_vs_spin}, we allow for 2000 likelihood evaluations rather than 1000. In most cases, the optimizer is quite stable, finding the value from PE within 2000 steps (often much fewer) given each initial point. Though the optimizer occasionally fails to find the correct value within 2000 iterations, the correct value agreeing with PE is always found at least once for every five initializations, and often up to five out of five times. In Fig.~\ref{fig:lnL_by_octant}, we show the results of initializing 50 times in the $(-1,\,0)$ and $(1,\,0)$ octants for the $M=10^7M_\odot,\;q=3,\;\iota=\pi/3$ event with $\rho_{\rm res}$ values shown on the left in Fig.~\ref{fig:rho_res_dlo_vs_pe}. A maximum log-likelihood value near the one found in a full PE is found at least $\sim\!40$\% of the time, suggesting that one can generally take the highest value from five initializations, and it will likely be a correct value. The top panel of Fig.~\ref{fig:lnL_by_octant} shows the distribution of $\ln \mathcal{L}_{\rm max}$ values over 50 iterations. In the bottom row, we show more clearly the particular $\ln \mathcal{L}_{\rm max}$ values found each time and how they compare to the value found in PE (red line). Note that the $\ln \mathcal{L}_{\rm max}$ values found by the optimizer can even be slightly higher than the maximum found in PE, given the finite number and spacing of samples from any given PE run. 
Given the high frequency with which the optimizer finds the same peak as PE, we find that likelihood optimization with dual annealing can be a powerful tool in rapidly finding the maximum likelihood parameters.  

\begin{table}[htbp]
\caption{\label{tbl:1e8} 
  Values of $\ln \mathcal{L}_{\rm max}$ as found in different octants of the sky by the direct likelihood optimization (``DLO'') method and PE for a massive event with $M=10^8\,M_\odot$. The difference between octants is not significant, in the sense that it cannot be distinguished from changes in the log-likelihood that will arise due to variations in the noise realization.} 
\begin{tabularx}{\linewidth}{X X X}
 \toprule
  \midrule
   \multicolumn{3}{c}{$M=10^8\,M_\odot,\;q=3,\;\iota=\pi/3$} \\
 \midrule
 $(s,\;n)$ & $\ln \mathcal{L}_{\rm max,\;DLO}$ & $\ln \mathcal{L}_{\rm max,\;PE}$ \\
 \midrule
 (-1, 0) & -41.15 & -41.79   \\
 (-1, 1) & -43.56 & -44.17   \\
 (-1, 2) & -40.73 & -41.43   \\
 (-1, 3) & -37.98 & -38.35   \\
 \textbf{(1, 0)} & \textbf{-39.46} & \textbf{-39.62}   \\
 (1, 1)  & -39.14 & -39.47   \\
 (1, 2)  & -42.66 & -42.93   \\
 (1, 3)  & -42.75 & -43.61   \\
 \midrule
  \bottomrule
\end{tabularx}
\end{table}

Unfortunately, as one continues increasing the total mass of the MBHB, the $\ln \mathcal{L}_{\rm max}$ values between different octants becomes nearly indistinguishable. This can be seen in Table~\ref{tbl:1e8}, where we show the $\ln \mathcal{L}_{\rm max}$ values as determined by PE and likelihood optimization in different octants. Here,  the values change very little from one region of the sky to another. In fact, the difference in $\ln \mathcal{L}_{\rm max}$ is significantly smaller than the expected change in log-likelihood due to random noise realizations: see Eq.~\eqref{eqn:logL_criterion}. For such massive binaries ($M\gtrsim10^8\,M_\odot$), it will be practically impossible to confidently localize the source to a certain octant of the sky using LISA, although the width of the sky localization posteriors within each octant may still be small. Note that in this highest mass range, to find the agreement with PE shown in Table~\ref{tbl:1e8}, we had to run the optimizer for an extra 1000 iterations. %
At this point, when the number of likelihood evaluations gets closer to rivaling the number for a full PE run, one could argue that the likelihood optimization begins to lose its advantage. In particular, if the posterior samples are distributed fairly equally between different octants (as suggested by Table~\ref{tbl:1e8}), it becomes all the more important to have a full PE run. Then one can try to compare the probability between different octants, since it is no longer simple to identify a single highest-probability region.

Finally, we note that for a $M=10^7\,M_\odot,\;q=1.1,\;\iota=\pi/12$ event,
the improved likelihood optimization scheme presented in this paper returns many different answers upon being initialized several times; it even occasionally finds a \emph{higher} $\ln \mathcal{L}_{\rm  max}$ than the parallel-tempered PE. In fact, the likelihood optimization also finds a higher peak than PE for the MBHB with $M=10^6\,M_\odot,\,q=1.1,\,\iota=\pi/12$ shown in Fig.~6 of~\citetalias{Yi:2025pxe} (i.e., for a system with all the same parameters except one tenth the total mass).
We therefore see that for very high-SNR, short signals for which it can be difficult to determine whether PE has converged, it can be useful to run the likelihood optimization method discussed here to help ascertain whether the highest peak has truly been found. 

The improved likelihood optimization method outlined in this section will be useful in future studies of systematic errors, and in particular, in determining which additional modes beyond the ones modeled in current state-of-the-art waveform approximants will ultimately be necessary for MBHB analysis. Our findings in Sec.~\ref{sec:spin_dependence} suggest that to accurately predict the necessity of, e.g., the $\ell\geq6$ modes, one will need to survey a broad parameter space of not only total mass, mass ratio, and inclination, but also progenitor spins. Such a large parameter space will be very expensive and time-consuming to survey with traditional Bayesian analyses. By using the likelihood optimization tool described here, we can more quickly and efficiently provide waveform developers with an understanding of which higher modes will eventually be necessary to model in order to perform unbiased parameter estimation on LISA sources. 

\section{Conclusions}\label{sec:conclusions}

In this work, we have extended the findings of~\citetalias{Yi:2025pxe} by considering the systematic biases due to neglected higher-order modes for more massive systems than were originally considered, as well as for MBHBs with nonzero progenitor spins. For MBHBs with $M\gtrsim5\times10^6\,M_\odot$ and $q\gtrsim5$, the (2,\,2) mode is not necessarily the most dominant, causing higher-order modes to become particularly important in performing accurate PE. We improved the likelihood optimization method introduced in~\citetalias{Yi:2025pxe} by implementing global optimization tools, reparametrizing the problem, and using Fisher-informed priors. This allowed us to explore the bias due to a neglected higher-order mode as a function of the progenitor spins. 

Importantly, we see in Fig.~\ref{fig:bias_vs_spin} that a significant fraction of MBHBs with total mass $M=10^6\,M_\odot$ and various combinations of $q,\,\iota,$ and spins will be biased beneath $z\sim2.5$. While predicted rates of MBHBs in LISA's band vary considerably, one can use the models in, e.g., Ref.~\cite{Barausse:2023yrx} as a benchmark. According to these models, 
about $\sim6-22\%$ of detectable MBHBs (total SNR > 12) in a 4-year LISA mission would be at $z \leq2.5$ 
(where we quote numbers for the ``popIII-d (K+16),'' ``Q3-d (K+16),'' and ``Q3-nod (K+16)'' models). We can anticipate that about the same number of MBHB events could have significantly biased parameter estimation due to a neglected higher-order mode (though one should note, of course, that we have just seen the extent of systematic bias to vary significantly with the particular combinations of masses, spins, and extrinsic parameters). 

\begin{figure*}[t]
\centering 
\includegraphics[width=0.98\textwidth]{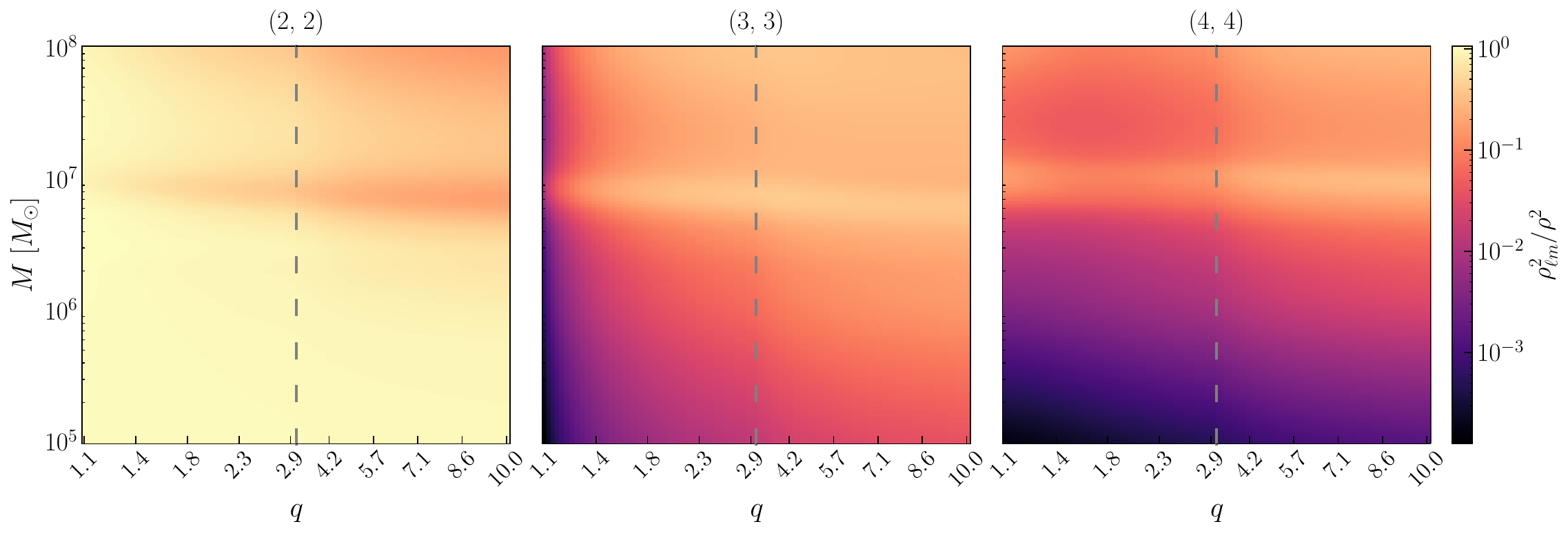} 
\includegraphics[width=0.66\textwidth]{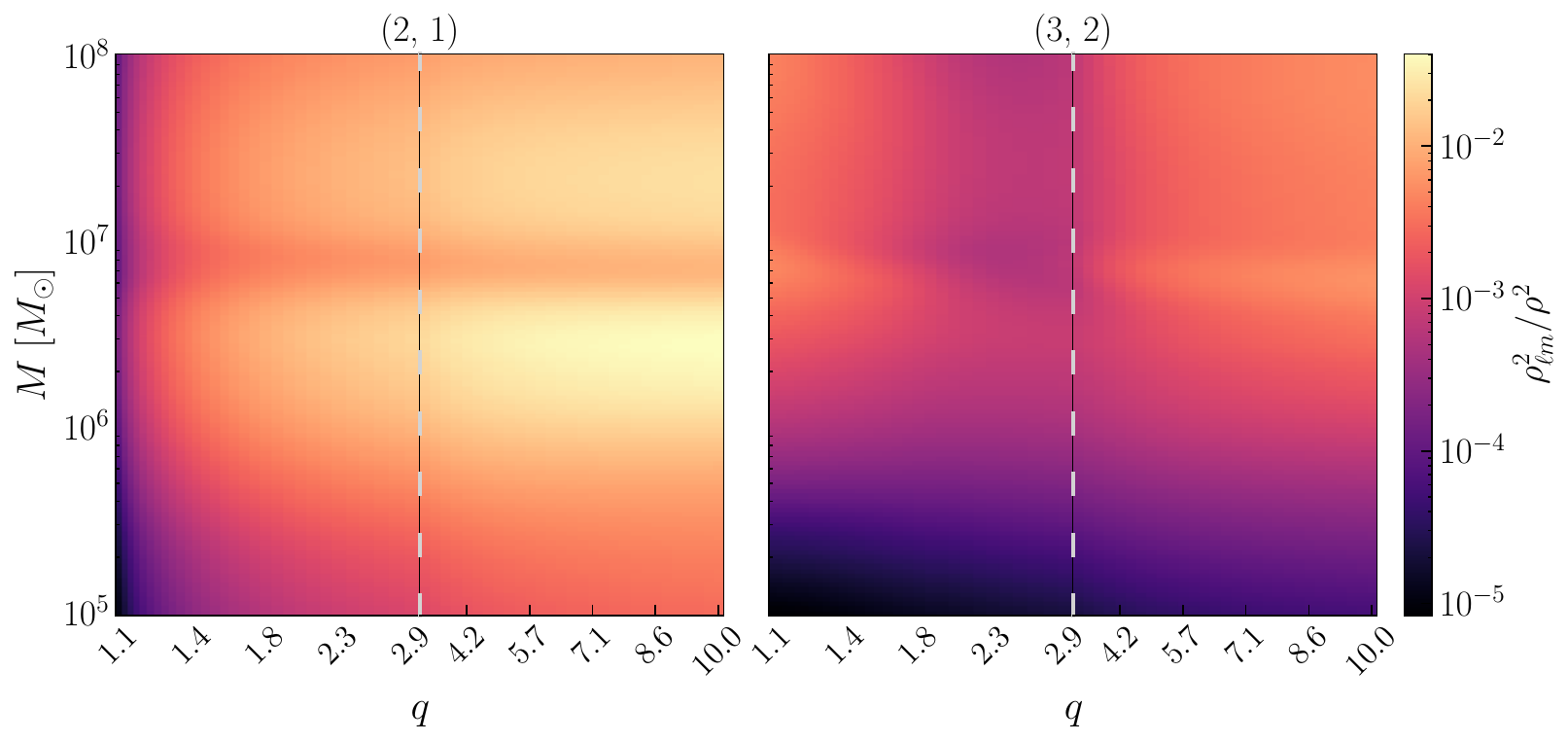} 
\caption{Fractional SNR squared contributed by each $(\ell,m)$ component of the signal, as a function of total mass and mass ratio. The inclination here is set to $\iota=\pi/3$, and component spins are set to zero. Note the different color scales used for the $\ell=m$ (top row) and $\ell\neq m$ modes (bottom row).}
\label{fig:frac_snr}
\end{figure*}

\begin{figure*}[tbp]
\centering
\includegraphics[width=0.9\textwidth]{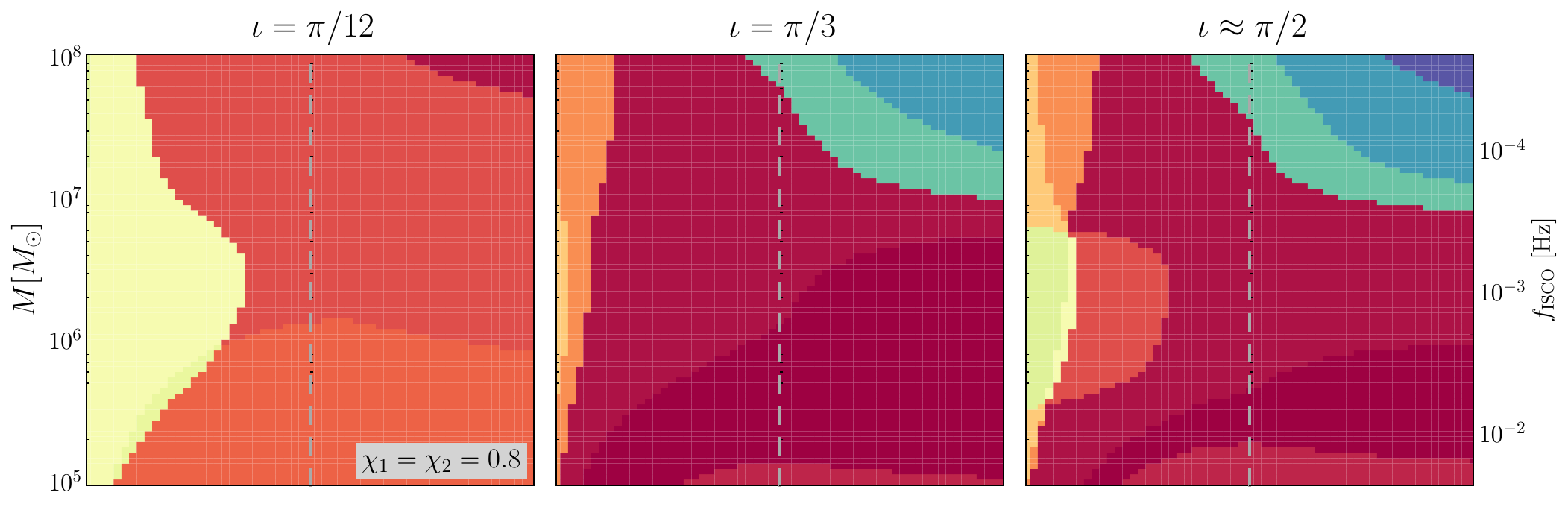} 
\includegraphics[width=0.9\textwidth]{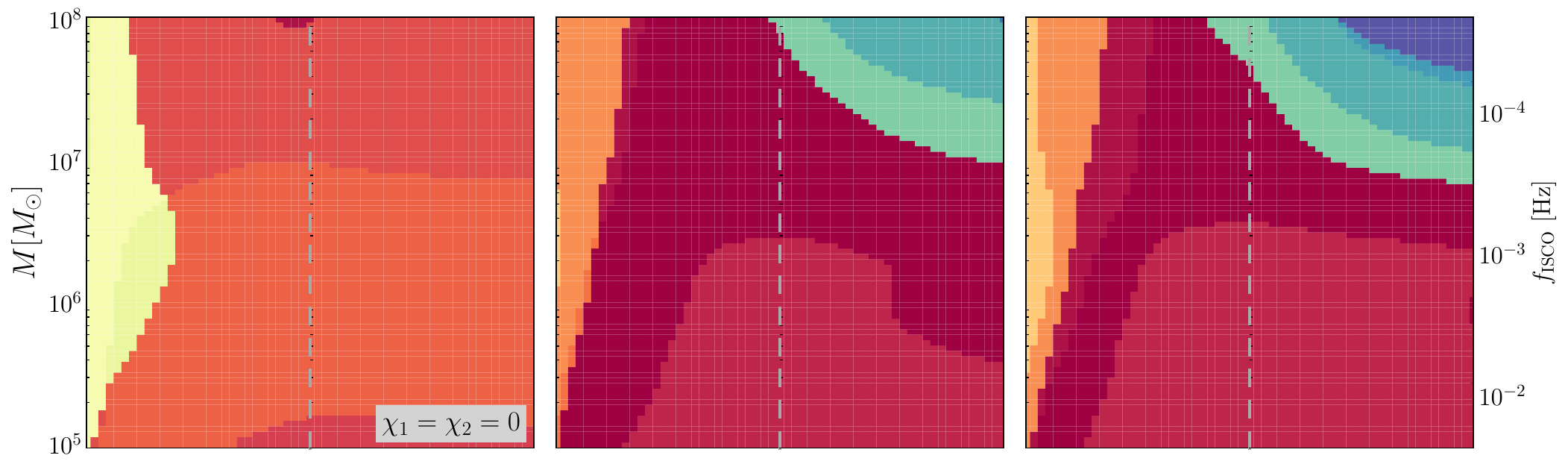} 
\includegraphics[width=0.9\textwidth]{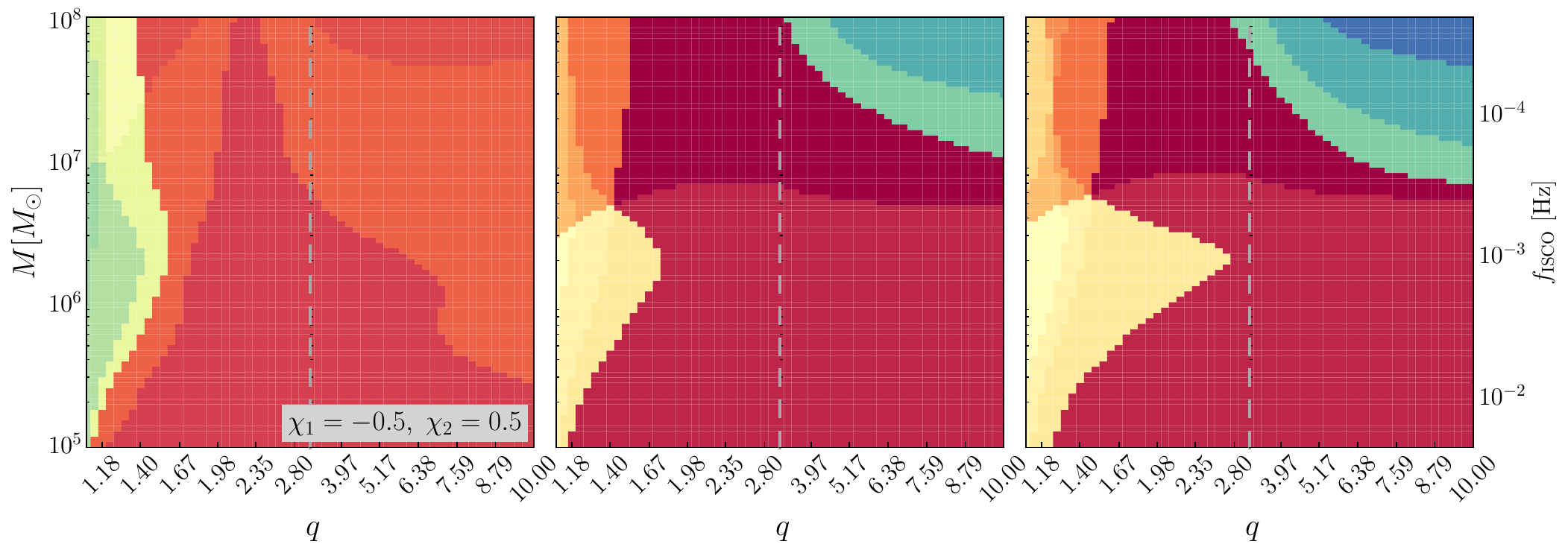} 
\hspace{12cm} \includegraphics[width=0.9\textwidth]{images/legend_distinct.pdf} 
\caption{Same as Fig.~\ref{fig:mode_by_mode}, but without the confusion noise from galactic binaries.}
\label{fig:mode_by_mode_noWD}
\end{figure*}

\begin{figure*}[tbp]
\centering
\includegraphics[width=0.98\textwidth]{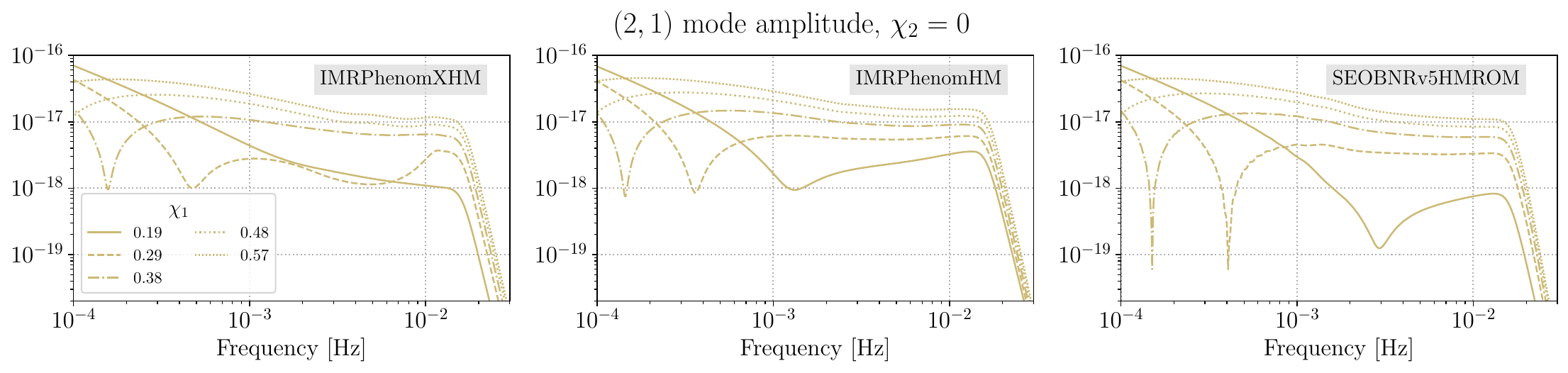} 
\includegraphics[width=0.98\textwidth]{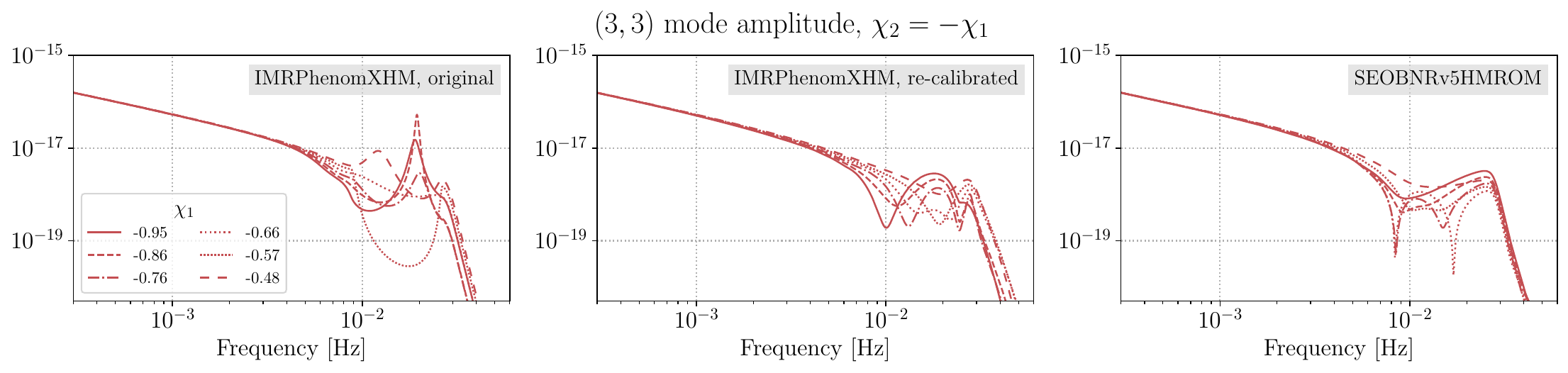} 
\caption{In the top panel, we show the $(2,\,1)$ mode amplitude in regions near the ``spikes'' shown in Fig.~\ref{fig:bias_vs_spin}. These ``spikes'' correspond to regions of parameter space where there are dips in the $(2,\,1)$ mode amplitude as shown. In addition to understanding this behavior from PN theory (see the discussion in Sec.~\ref{sec:spin_dependence}), we note that multiple different waveform approximants seem to capture this behavior (shown here are the \textsc{IMRPhenomXHM}, \textsc{IMRPhenomHM}, and \textsc{SEOBNRv5HM\_ROM} waveforms). 
In the bottom panel, we show a plot of $(3,\,3)$ mode amplitudes at low $\chi_1$ for the $q=1.1$ case shown in Fig.~\ref{fig:bias_vs_spin}, as modeled by the \textsc{IMRPhenomXHM} and \textsc{SEOBNRv5HM\_ROM} waveforms. 
Throughout this work, we use the original release of \textsc{IMRPhenomXHM} (``version \texttt{122019}'', shown in the left panel). The recalibrated version (``version \texttt{122022}'', center panel) looks more similar to the \textsc{SEOBNRv5HM\_ROM} model and does not have the unexpected features shown in the left panel. 
}
\label{fig:wf_amps}
\end{figure*}

Our improved likelihood optimization method also allowed us to rapidly find the octant of the sky containing the highest $\ln \mathcal{L}_{\rm max}$ in situations where the brevity of the signal causes multiple modes to appear in the posteriors of the sky localization parameters. For the most massive systems considered ($M\gtrsim10^8\,M_\odot$), the degeneracy becomes quite severe between different regions of the sky, and one can no longer confidently determine which octant contains the highest peak when accounting for the fluctuations in log-likelihood due to random noise realizations. 

Although we have expanded the work of~\citetalias{Yi:2025pxe} in this present work, much remains to be done. For instance, we have still restricted our analysis thus far to nonprecessing, quasicircular binaries. Eventually, we will need to understand and mitigate the systematic biases due to not only neglected higher modes, but also neglected precession and eccentricity as well. Moreover, as noted in~\citetalias{Yi:2025pxe}, it will ultimately be essential to predict the bias due to neglecting even higher-order modes than are currently unavailable in many waveform approximants (e.g., the $\ell\geq6$ modes, as well as additional $m\neq\ell$ modes for $\ell>2$). Our findings in this paper suggest that in making these predictions, it will be crucial to account for the effect of progenitor spins in making a given higher-order mode relevant in a given region of parameter space. Neglecting such spin effects could cause a significant over- or under-estimation of bias due to a neglected higher mode (see Fig.~\ref{fig:bias_vs_spin}). Given the vast parameter space that will thus need to be covered when investigating the necessity of additional higher modes, we propose the use of our improved likelihood optimization tool (Sec.~\ref{sec:confirm_NM_higher_mass}) to explore this question in a computationally efficient manner.

Finally, using inaccurate waveforms for loud MBHBs could potentially result in significant residuals in the data stream that can impact the LISA global fit effort~\cite{Littenberg:2020bxy,Littenberg:2023xpl,Katz:2024oqg,Deng:2025wgk,Khukhlaev:2025xiz}, especially in inferring quieter sources.
We hope that such studies can benefit from the likelihood optimization method utilized in~\citetalias{Yi:2025pxe} and further developed here. 

\begin{acknowledgments}

We thank Chantal Pitte, Manuel Piarulli, and Aasim Jan for very useful discussions and comments on the draft. S.Y.  is supported by the NSF Graduate Research Fellowship Program under Grant No. DGE2139757.
S.Y., F.I., D.W, and E.B. are supported by NSF Grants No.~AST-2307146, No.~PHY-2513337, No.~PHY-090003, and No.~PHY-20043, by NASA Grant No.~21-ATP21-0010, by John Templeton Foundation Grant No.~62840, by the Simons Foundation [MPS-SIP-00001698, E.B.], by the Simons Foundation International [SFI-MPS-BH-00012593-02], and by Italian Ministry of Foreign Affairs and International Cooperation Grant No.~PGR01167. The work of F.I. is supported by a Miller Postdoctoral Fellowship. S.M. acknowledges support from the French space agency CNES in the framework of LISA. R.S.C. acknowledges support from the European Union’s Horizon ERC Synergy Grant ``Making Sense of the Unexpected in the Gravitational-Wave Sky''(Grant No.~GWSky-101167314), and from the PRIN 2022 grant
``GUVIRP -- Gravity tests in the UltraViolet and InfraRed with Pulsar timing''.
N.Y.~acknowledges support from the Simons Foundation through Award No. 896696, the Simons Foundation International through Award No. SFI-MPS-BH-00012593-01, the NSF through Grants No. PHY-2207650 and PHY-25-12423, and NASA through Grant No. 80NSSC22K0806.
This work was carried out at the Advanced Research Computing at Hopkins (ARCH) core facility (\url{rockfish.jhu.edu}), which is supported by the NSF Grant No.~OAC-1920103. 
\end{acknowledgments}

\appendix 

\section{Further study of mode-by-mode SNR ordering}\label{app:frac_snr}

In Fig.~\ref{fig:frac_snr}, we plot the relative contribution of each mode compared to the total SNR of the signal. Because of the closer connection to the log-likelihood and bias, we show the fractional contribution of SNR squared ($\rho^2$) for each mode (see Eq.~\eqref{eqn:inner_product}), rather than just the SNR. Comparing with Fig.~\ref{fig:mode_by_mode}, we see that even in regions where the (2,\,2) mode is technically loudest (red, orange, or yellow in Fig.~\ref{fig:mode_by_mode}), this mode might really only be contributing a few tens of percent of the total SNR squared (i.e., some $\sim\!60\%$ of the SNR). From these figures, it is also clear how a lower relative contribution of the (2,\,2) mode to the total signal is directly correlated with a higher relative contribution of the subdominant modes, particularly the (3,\,3) and (4,\,4). The fractional SNR squared for the (2,\,2) mode can be larger than one, due to the fact that cross terms can contribute negatively to the total SNR (see Sec.~IV of Ref.~\cite{Pitte:2023ltw}).

We also note that the (3,\,2) mode never contributes more than a few percent of the SNR squared. Nevertheless, we have seen throughout the main text as well as in~\citetalias{Yi:2025pxe} that neglecting even this one mode can result in significant biases on the inferred MBHB parameters. As mentioned in the main text, this is at least partly because there can be significant contribution from this mode via its cross product with other modes. Within the same parameter space of Fig.~\ref{fig:frac_snr}, the cross term $\left(h_{22}|h_{32}\right)$ can contribute up to 5\% (in absolute value) of the total squared SNR. Other significant cross terms include the $\left(h_{22}|h_{33}\right)$ and $\left(h_{33}|h_{44}\right)$, which each contribute up to 10\% of the total squared SNR.

Lastly, we comment further on the impact of the galactic background on the relative importance of different modes. In Fig.~\ref{fig:mode_by_mode_noWD}, we reproduce the SNR hierarchy plot shown in Fig.~\ref{fig:mode_by_mode}, but this time removing the confusion noise from galactic binaries. By comparing the two figures, it is clear that the features along the horizontal around $\sim10^7\,M_\odot$ in Fig.~\ref{fig:mode_by_mode} are due to the presence of the confusion noise. Without this feature, the hierarchy between the (2,\,2), (3,\,3), and (4,\,4) modes is much smoother as the total signal is pushed to lower and lower frequencies due to the increase in total mass of the system (i.e., moving toward the top of the plots). 

\section{Comparison of mode amplitudes as modeled by different waveform families}\label{app:compare_wfs}

In this appendix, we illustrate the behavior of waveform amplitudes at certain BBH parameter values which lead to various features alluded to in the main text. 

\subsection{\texorpdfstring{$(2,\,1)$}{(2, 1)} mode amplitudes in the near-equal-mass case at low spins}

In Sec.~\ref{sec:spin_dependence}, we find ``spikes'' in the amount of systematic bias due to neglecting the $(3,\,2)$ mode in the $q=1.1,\,\iota=\pi/3$ case. We explain this behavior with the PN structure of the $(2,\,1)$ mode, noting the competition between LO and NLO contributions to this mode's amplitude depending on the values of the spin. This competition can be seen in the top panel of Fig.~\ref{fig:wf_amps}, where there are dips as a function of the frequency for various spin combinations shown in the $q=1.1$ panel of Fig.~\ref{fig:bias_vs_spin}. Notably, this behavior is captured by several different waveform approximants; we show the features as modeled by the \textsc{IMRPhenomHM} and \textsc{SEOBNRv5HM\_ROM} families, in addition to the \textsc{IMRPhenomXHM} family used throughout the main text. In all three cases, we see similar ``peaks'' in the systematic bias around these values of the spins. Importantly, this behavior can also be observed in NR simulations, further verifying its physical origin~\cite{Cotesta:2018fcv,Nagar:2020pcj,Estelles:2020twz,Scheel:2025jct}.

\begin{figure*}[htbp]
\centering
\begin{tabular}{l@{\hskip .4cm}l}
    \includegraphics[width=0.45\textwidth]{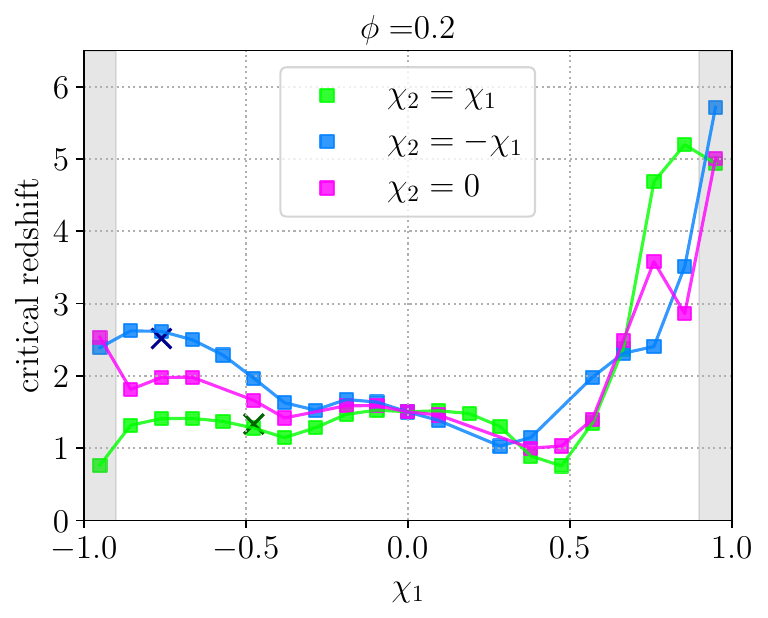} & \includegraphics[width=0.45\textwidth]{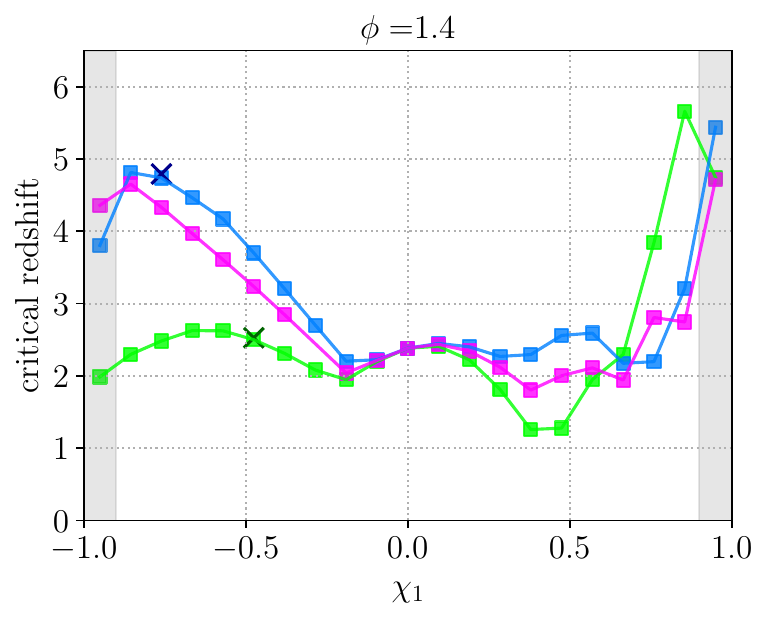} \\
    \includegraphics[width=0.45\textwidth]{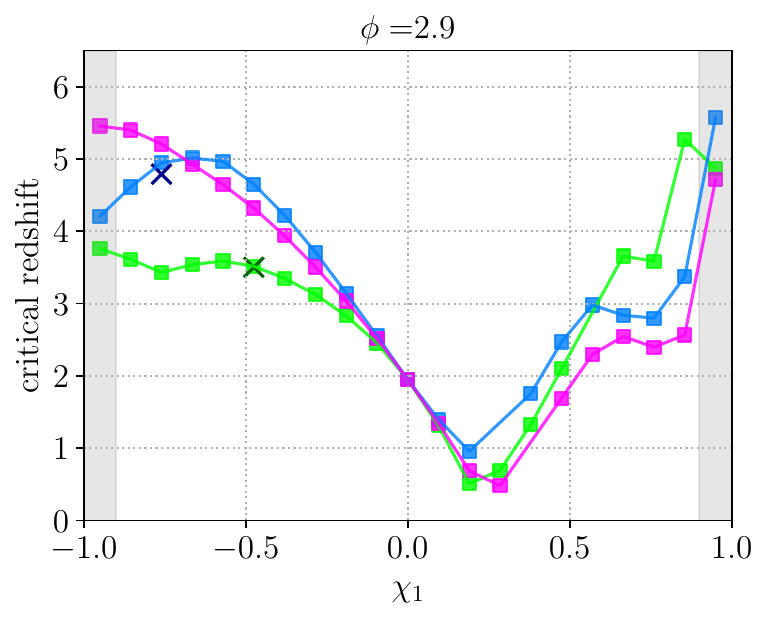}  & \includegraphics[width=0.45\textwidth]{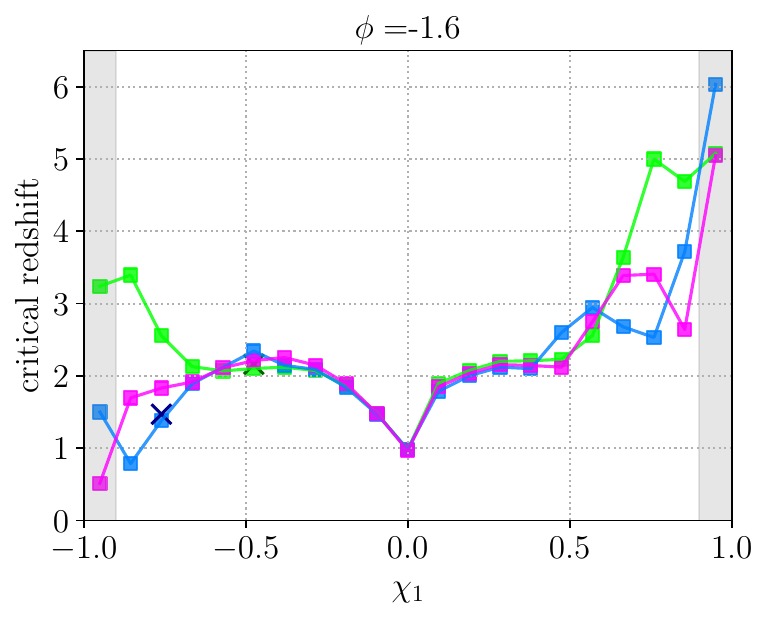} 
\end{tabular}
\caption{Impact of the phase parameter $\phi$ on the degree of bias observed. The top left panel is equivalent to the top left panel of Fig.~\ref{fig:bias_vs_spin}. In the other panels, we modify the parameter $\phi$ with respect to the original configuration. There is a moderate impact on the degree of bias observed.}
\label{fig:bias_vs_spin_var_phi}
\end{figure*}

\subsection{\texorpdfstring{$(3,\,3)$}{(3, 3)} mode amplitudes in the near-equal-mass case at extreme values of \texorpdfstring{$\chi_1$}{chi1} and \texorpdfstring{$\chi_2$}{chi2}}

In Fig.~\ref{fig:bias_vs_spin}, we exclude points to the left of $\chi_1=-0.38$ for the blue $\chi_2=-\chi_1$ case. Upon noticing that the optimizer struggled somewhat more in this region compared to other regions, we examined the waveforms in this range and found the behavior plotted in the left panel of Fig.~\ref{fig:wf_amps} for the amplitude of the $(3,\,3)$ mode. These amplitudes look notably different as modeled by other waveform families, including \textsc{SEOBNRv5HM\_ROM} (right panel of Fig.~\ref{fig:wf_amps}). As we do not see a straightforward way to explain this behavior, we choose not to plot estimates for the systematic errors at these points. We note that the later, recalibrated version of \textsc{IMRPhenomXHM} (``version \texttt{122022}'') is more similar to the \textsc{SEOBNRv5HM\_ROM} model, but it is the original version (``version \texttt{122019}'') that is directly implemented within \texttt{lisabeta} and used throughout this paper. For comparison, we also plot the recalibrated version in the middle panel of Fig.~\ref{fig:wf_amps}.

\section{Dependence of biases on reference phase}\label{app:vary_phi}

In the main text, we have explored the impact of neglected higher-order modes as a function of the total mass, mass ratio, spins, and inclination of MBHB systems observed by LISA. In this section, we will further investigate the extent to which the severity of biases is affected by different reference phases. 

In Fig.~\ref{fig:bias_vs_spin_var_phi}, we plot the same results for the $M=10^6M_\odot,\;q=4,\;\iota=\pi/3$ binaries in the top row of Fig.~\ref{fig:bias_vs_spin} but with different values of the phase, $\phi$, defined as the azimuthal angle of the observer in the source frame. In the top left panel, we reproduce the top left panel of Fig.~\ref{fig:bias_vs_spin} for reference. Then, we select three additional values in the range $[-\pi, \,\pi]$ and redo the inference, holding the rest of the parameters fixed. 

Here we see that changing this phase parameter does indeed have a moderate impact on the degree of bias as a function of progenitor spins. Given the large differences at negative $\chi_1$ values, we perform several full Bayesian runs to validate the likelihood optimization results; the outcomes of these validation runs are marked by dark blue and green crosses. Given the good agreement between the likelihood optimization and full PE, we take the observed dependence on $\phi$ to be robust. 

The moderate impact we see is in large part due to the matter of cross-terms in the SNR mentioned in Sec.~\ref{sec:spin_mode_ordering}; different values of $\phi$ change the magnitude and even the sign of different products of $h_{\ell m}$'s, causing the degree of bias due to a neglected mode to change accordingly. We leave a more extensive analysis on the effect of $\phi$ and other extrinsic parameters to future work.

\bibliography{lisaHM2}

\end{document}